# Surprise Reduction and Nullification in Bayesian and Inverse Bayesian Inference under Ambiguous Prediction-Error Attribution

Shuji Shinohara[a,*], Daiki Morita[a], Yoshihiro Nakajima[b], Takeshi Takano[c], Masakazu Higuchi[d], Ung-il Chung[e], and Yukio-Pegio Gunji[f]

[a] School of Science and Engineering, Tokyo Denki University, Hatoyama, Saitama, Japan

[b] Graduate School of Economics, Osaka Metropolitan University, Osaka, Japan

[c] Department of Advanced Energy, Graduate School of Frontier Sciences, The University of Tokyo, Kashiwa, Chiba, Japan

[d] Department of Fundamental Engineering (Information, Electronics, and Optics Course), Utsunomiya University, Utsunomiya, Tochigi, Japan

[e] Department of Bioengineering, Graduate School of Engineering, The University of Tokyo, Tokyo, Japan

[f] Department of Intermedia Art and Science, School of Fundamental Science and Engineering, Waseda University, Tokyo, Japan

***Corresponding author:** Shuji Shinohara, School of Science and Engineering, Tokyo Denki

University, Hatoyama, Saitama, Japan. E-mail: s.shinohara@mail.dendai.ac.jp

ORCID iDs:

Shuji Shinohara: 0000-0001-8442-836X

Masakazu Higuchi: 0000-0003-0329-1141

Takeshi Takano: 0000-0002-0669-0175

Ung-il Chung: 0000-0002-4691-6394

## Acknowledgements

The authors have no acknowledgements to declare.

## Statements and Declarations

**Funding:** No funding was received for conducting this study.

**Competing interests:** The authors have no relevant financial or non-financial interests to disclose.

**Ethics approval:** Not applicable. This study involved computational simulations only and did not involve human participants, human data, or animals.

**Consent to participate:** Not applicable.

**Consent for publication:** Not applicable.

**Data availability:** No datasets were generated other than simulation outputs, which can be fully reproduced using the publicly available source code.

**Code availability:** The source code used for the simulations is publicly available on GitHub at

https://github.com/shinoharaken/BIBInference1.

**Author contributions:** Shuji Shinohara: Conceptualization, Formal analysis, Methodology, Software, Writing—original draft preparation. Daiki Morita: Conceptualization, Writing—review and editing. Yoshihiro Nakajima: Conceptualization, Writing—review and editing. Takeshi Takano: Writing—review and editing. Masakazu Higuchi: Writing—review and editing. Ung-il Chung: Writing—review and editing, Supervision. Yukio-Pegio Gunji: Writing—review and editing, Supervision.

**Use of generative AI:** During the preparation of this manuscript, the authors used ChatGPT (OpenAI) and Claude (Anthropic) to assist with language editing and the improvement of clarity and expression. All AI-assisted output was reviewed and revised by the authors, who take full responsibility for the content of the manuscript.

## Abstract

In non-stationary environments, prediction errors may signal environmental change or transient outliers, and adaptive systems must track such changes without overreacting to outliers. We distinguish surprise reduction, which updates beliefs to fit observations, from surprise nullification, which weakens constraints imposed by the predictive structure, and formalize both within Bayesian and inverse Bayesian (BIB) inference.

Belief and likelihood updates are derived from variational objectives sharing a nullification strength, determined endogenously by minimizing surprise under the candidate post-update predictive distribution. In the Gaussian case, nullification expands belief and likelihood variances by a common factor relative to standard Bayesian updating, leaving the ratio unchanged. BIB thus defers attribution of the prediction error, committing to neither latent-state change nor observation-process uncertainty. The nullification strength is carried over as a candidate and is maintained or released according to the predictive surprise of the next observation. In a mean estimation task with outliers and changepoints, no scanned parameter setting of a Sage–Husa-type adaptive Kalman filter, fixed-strength BIB variant, or belief-forgetting-only variant outperforms BIB in both changepoint tracking and post-outlier stability.

An oracle-informed reduced Bayesian model tracks changepoints better but is less stable after outliers. Although BIB maintains no explicit hypotheses about changepoints or outliers, it generates event-dependent dynamics. The learning rate increases after changepoints, whereas after outliers, nullification is released, and this increase is suppressed. Deferring attribution and letting subsequent observations differentiate the responses may constitute a principle of adaptive inference in non-stationary environments.

## Keywords

# 1 Introduction

Adaptive systems respond to environmental changes by continually resolving mismatches between the predictions of an internal model and the incoming observations. The free-energy principle casts perception and learning as the minimization of variational free energy (Friston 2010; Friston et al. 2017), whereas the predictive processing framework characterizes the brain as a hierarchical predictive system (Clark 2013). Within these frameworks, updating beliefs to make them consistent with observations is a fundamental inferential process whose canonical form is a standard Bayesian inference. In this study, we refer to this process of standard Bayesian updating as *surprise reduction*.

However, a large prediction error in a non-stationary environment may indicate a persistent environmental change, that is, a changepoint, or it may reflect a transient outlier. In the former case, the estimate should rapidly track the post-change state, whereas in the latter case, it should avoid being unduly influenced by a single observation. Indeed, humans dynamically adjust the extent to which newly acquired information affects subsequent learning and decision-making according to the environmental volatility, changepoint probability, and uncertainty in their estimates (Behrens et al. 2007; Nassar et al. 2010; McGuire et al. 2014).

Moreover, even for prediction errors of comparable magnitude, the learning rate increases in a changepoint context but decreases in an outlier context (Nassar et al. 2019). Thus, humans appear to regulate the updating of their internal models according to the magnitude of the prediction error as well as its inferred source.

Existing methods have addressed the problem of reconciling responsiveness to changepoints with stability against outliers in several directions. Bayesian inference with forgetting weakens the constraint imposed by past beliefs and thereby enhances responsiveness to environmental change (Kulhavý and Zarrop 1993; Kulhavý 1996). In contrast, robust filtering based on generalized Bayesian inference suppresses the influence of anomalous observations, thereby improving robustness to outliers (Duran-Martin et al. 2024). In changepoint models, the weight assigned to an observation during belief updating is adjusted according to the changepoint probability (Nassar et al. 2010). Normative models that treat changepoint and outlier conditions separately have also been proposed (Nassar et al. 2019).

Classical adaptive Kalman filtering provides methods for adapting processes and/or observation noise statistics in response to changing environmental conditions (Jazwinski 1969; Sage and Husa 1969). Hierarchical Bayesian models that jointly estimate latent-state volatility and observation stochasticity have also been proposed (Piray and Daw 2021), and this simultaneous inference in humans has subsequently been examined (Piray and Daw 2024).

Another framework provides a unified description of diverse Bayesian online learning methods for non-stationary environments in terms of design components such as the observation model, auxiliary process, and conditional prior (Duran-Martin et al. 2025). However, in these approaches, the types of environmental changes or observation characteristics to be explicitly represented are specified in advance through the model structure.

Relative to standard Bayesian updating, the responses to a large prediction error can be organized in two principal directions. When the prediction error is attributed to a change in the latent state, weakening the constraint imposed by past beliefs, thereby incorporating the observation more strongly than under standard Bayesian updating, is an effective response. We refer to this as *belief relaxation*. In contrast, when the prediction error is attributed to an anomaly or uncertainty in the observation process, an effective response is to reduce the certainty of the correspondence between the observation and latent states, thereby suppressing the influence of the observation on belief updating. The operation of weakening the constraint imposed by the likelihood is referred to as *likelihood relaxation*. The former corresponds to a response to a change on the state side, whereas the latter corresponds to a response to an anomaly on the observation side. Existing methods do not necessarily classify the cause definitively; some represent state-side and observation-side factors separately and estimate their respective contributions.

In contrast, the present study considers a third mode of operation that considers unresolved attribution as the starting point. From a single large prediction error, it is generally difficult to determine whether the error arises from a change in the latent state or an anomaly during the observation process. Therefore, we weaken both the constraint imposed by past beliefs and that imposed by likelihood simultaneously. This temporarily relaxes the entire predictive structure, consisting of the belief and observation models, and leaves the attribution of the prediction error open to subsequent observations. We refer to this response as *surprise nullification*. Therefore, nullification is an operation that temporarily withholds the preferential attribution of a prediction error to either the state side or observation side.

In this study, surprise nullification is realized through Bayesian and inverse Bayesian (BIB) inference, which updates the belief distribution as well as likelihood structure (Gunji et al. 2017). Shinohara et al. (2020a, b) proposed discrete update rules that extend Bayesian inference by incorporating symmetry bias (Shinohara et al. 2007), the tendency to infer "if q then p" from "if p then q." Shinohara et al. (2026) subsequently formulated a continuous BIB inference in which BIB updates are coupled through a symmetry-bias strength β. In that formulation, however, β was supplied externally and held fixed. In the present study, we reinterpret β as a nullification strength that simultaneously weakens the belief and likelihood constraints and determine it endogenously to minimize the surprise of the current observation under the candidate post-update predictive distribution. $\beta = 0$ corresponds to standard Bayesian updating, whereas $\beta > 0$ weakens the constraints on the belief and likelihood sides with a common strength. Thus, β does not merely denote an amount of belief forgetting or observation suppression; rather, it represents the degree to which the constraints imposed by the entire predictive structure are weakened relative to standard Bayesian updating. However, the value of β determined from the current observation is not applied immediately to the update for that same observation. Instead, it is retained as a candidate for the next time step.

We examined the effects of nullification and delayed application in a non-stationary mean estimation task in which transient outliers and abrupt environmental changes were intermixed. For comparison, we used a Sage–Husa-type adaptive Kalman filter (SH) and an oracle-informed reduced Bayesian model (RB), which was constructed by combining the reduced Bayesian model proposed by Nassar et al. (2010) with the changepoint/oddball framework developed by Nassar et al. (2019).

The simulation results indicate that, at least in the present task setting, BIB can temporarily defer the attribution of a prediction error and subsequently differentiate its responses in terms of responsiveness and stability without using prior information about causal hypotheses or event rates.

## 2 Model

In this section, we first define standard Bayesian updating as the baseline surprise reduction process. Relative to this baseline, we then derive two update rules from separate variational objectives: the B-step, which weakens the constraint imposed by past beliefs (belief relaxation), and the L-step, which weakens the constraint imposed by the likelihood on the observation–state correspondence (likelihood relaxation). Finally, we formalize surprise nullification as the simultaneous application of the B-step and L-step under a shared nullification strength $\beta$.

Detailed derivations of the model equations presented in this section are provided in the Appendix.

## 2.1 Variational formulation of surprise reduction and surprise nullification

### 2.1.1 Surprise reduction: standard Bayesian updating

Let $s$ denote the latent state and $o_t$ denote the observations obtained at time $t$. The inferring agent represents its belief in the latent state before observing $o_t$ using probability distribution $q_t(s)$. The likelihood of obtaining observation $o_t$ given latent state $s$ is denoted by $l_t(o_t \mid s)$.

The predictive distribution constructed from the pre-update belief and the likelihood is

$$p_t(o_t) = \int q_t(s) l_t(o_t | s) ds. \tag{1}$$

In this study, we refer to the process of updating beliefs to make them consistent with observations as surprise reduction, and we take standard Bayesian updating as its baseline process.

Standard Bayesian updating can be expressed as a variational problem that reduces the mismatch with the current observation, while penalizing the deviation from the pre-update belief. For a candidate belief $q(s)$, we define

$$F[q] = -\int q(s) \log l_t(o_t | s) ds + D_{KL}(q(s) \| q_t(s)). \tag{2}$$

This objective function can be rewritten as $F[q] = -\log p_t(o_t) + D_{KL}(q(s) \| p_t(s | o_t))$, where $F[q] \geq -\log p_t(o_t)$, and the minimum is attained when

$$q_{t+1}(s) = p_t(s|o_t) = \frac{q_t(s)l_t(o_t|s)}{p_t(o_t)}. \tag{3}$$

Thus, standard Bayesian updating can be expressed as a process that minimizes the variational free energy, attains its lower bound given by the predictive surprise $-\log p_t(o_t)$, thereby forming a belief consistent with the current observation.

Taking standard Bayesian updating as the baseline, we next formalize the operation that weakens the constraint imposed by the pre-update belief as the B-step and the operation that weakens the constraint imposed by the likelihood on the observation–state correspondence as the L-step.

### 2.1.2 B-step: belief relaxation

In B-step, the constraint imposed by the pre-update belief $q_t(s)$ is relaxed, allowing the current observation to exert a stronger influence than that under standard Bayesian updating.

For a candidate belief $q(s)$ and a relaxation strength $\beta \in [0,1]$, we define the following objective function:

$$F_B[q;\beta] = -\int q(s)\log l_t(o_t|s)ds + (1-\beta)D_{KL}[q \,||\, q_t] + \beta D_{KL}[q \,||\, u], \tag{4}$$

where $u(s)$ is the reference distribution for belief forgetting. The first term represents the cost of mismatch with the current observations. The remaining two terms form a convex combination: the second preserves the pre-update belief with weight $1-\beta$, whereas the third promotes movement toward the reference distribution with weight $\beta$.

Minimizing this objective function yields

$$\tilde{q}_{t+1}(s;\beta) \propto q_t(s)^{1-\beta} u(s)^{\beta} l_t(o_t|s). \tag{5}$$

In the Gaussian setting adopted below, $u(s)$ can be interpreted as the limit of a reference distribution with an arbitrarily large variance. More precisely, let $u_\Lambda(s)$ be a Gaussian reference distribution with variance $\Lambda$, and consider the limit $\Lambda \to \infty$. In this limit, the $s$-dependent part of $u_\Lambda(s)^\beta$ vanishes, while its remaining contribution is independent of $s$ and can therefore be absorbed into the normalization constant. Hence,

$$\tilde{q}_{t+1}(s;\beta) \propto q_t(s)^{1-\beta} l_t(o_t|s). \tag{6}$$

When $\beta = 0$, this update reduces to standard Bayesian updating,

$$\tilde{q}_{t+1}(s;0) = p_t(s|o_t). \tag{7}$$

For $\beta > 0$, the contribution of the pre-update belief is reduced, thereby increasing the relative influence of the current observation. Therefore, the B-step enhances the responsiveness to the observation relative to standard Bayesian updating by relaxing the constraint imposed by the pre-update belief.

### 2.1.3 L-step: likelihood relaxation

In the L-step, the certainty of the likelihood $l_t(o \mid s)$, which specifies the relationship between the latent state and the observation, is relaxed.

Given the pre-update belief $q_t(s)$, we define the following objective function for a candidate likelihood $l(o \mid s)$:

$$F_L[l;\beta] = \mathbb{E}_{q_t(s)}\left[D_{KL}\left(l(\cdot|s) \,||\, l_t(\cdot|s)\right)\right] + \beta \mathbb{E}_{q_t(s)}\mathbb{E}_{l(\cdot|s)}\left[\log p_t(o)\right], \tag{8}$$

or equivalently,

$$F_L[l;\beta] = \mathbb{E}_{q_t(s)}\mathbb{E}_{l(o|s)}\left[\log\frac{l(o|s)}{l_t(o|s)} + \beta \log p_t(o)\right]. \tag{9}$$

The first term is conservative and penalizes deviations from the pre-update likelihood. The second term refers to the current predictive distribution $p_t(o)$ and acts to weaken the existing observation–state correspondence.

Minimizing this objective function for each state $s$ yields

$$\tilde{l}_{t+1}(o|s;\beta) = \frac{l_t(o|s)\, p_t(o)^{-\beta}}{Z_t(s;\beta)}, \tag{10}$$

where $Z_t(s;\beta) = \int l_t(o|s)\, p_t(o)^{-\beta}\, do$ is the normalization constant, which is assumed to be finite.

If we hold the belief $q_t(s)$ fixed and consider the marginal quantity before normalizing the likelihood, then

$$\int q_t(s)\, l_t(o|s)\, p_t(o)^{-\beta}\, ds = p_t(o)^{1-\beta}. \tag{11}$$

When $\beta = 0$, the likelihood remains unchanged. For $\beta > 0$, the L-step compresses the contrast in the unnormalized predictive mass. In the fixed-center Gaussian family considered below, this effect is realized as an expansion of the likelihood variance, thereby reducing the amount of state information carried by the observation.

### 2.1.4 Integration of the B-step and L-step: surprise nullification

The B-step facilitates incorporation of the current observation by partially forgetting the pre-update belief, whereas the L-step relaxes the certainty of the likelihood, thereby reducing the state information that the observation carries in the subsequent predictive structure. In the BIB nullification candidate, these two operations are combined under a shared strength $\beta$. Specifically, we relax the constraint on the belief side as follows:

$$\tilde{q}_{t+1}(s;\beta) \propto q_t(s)^{1-\beta}\, l_t(o_t|s), \tag{12}$$

while, simultaneously, the constraint on the likelihood side is relaxed through

$$\tilde{l}_{t+1}(o|s;\beta) \propto l_t(o|s)\, p_t(o)^{-\beta}. \tag{13}$$

When $\beta = 0$, the B-step reduces to standard Bayesian updating, while the L-step leaves the likelihood unchanged. For $\beta > 0$, the B-step, which enhances responsiveness to the observation, and the L-step, which weakens the observation–state correspondence, operate simultaneously with a common strength. In this study, we refer to this simultaneous relaxation of the two constraints as surprise nullification. The candidate likelihood $\tilde{l}_{t+1}(\cdot\,|s;\beta)$ enters the predictive distribution only from the next time step onward. Within the current update, the influence of the observation is governed by $q_t$ and $l_t$ alone. Therefore, the increased incorporation of the observation produced by the B-step is not offset by the L-step at the same time step. The effect of likelihood relaxation appears in the uncertainty of the subsequent prediction, rather than in the current belief update.

### 2.1.5 Endogenous determination of the nullification strength

For a fixed candidate belief $q$ or a fixed candidate likelihood $l$, both $F_B[q;\beta]$ and $F_L[l;\beta]$ are affine functions of $\beta$. Consequently, the functions are partially minimized.

$$G_B(\beta) = \min_q F_B[q;\beta], \qquad G_L(\beta) = \min_l F_L[l;\beta] \tag{14}$$

are concave in $\beta$, because they are lower envelopes of affine functions, and their minima over the interval $0 \leq \beta \leq 1$ are therefore attained at the endpoints.

Minimizing either objective with respect to $\beta$ therefore admits an endpoint minimizer and, in general, does not uniquely determine an interior value of $\beta$. Moreover, there is no guarantee that the two objectives select the same endpoint. Determining $\beta$ by a criterion external to $F_B$ and $F_L$ is therefore structurally necessary rather than a matter of design convenience.

Therefore, we determined the common nullification strength based on the candidate predictive distribution obtained by applying the B-step and L-step simultaneously. From the candidate belief and candidate likelihood constructed under nullification strength $\beta$, we define

$$\tilde{p}_{t+1}(o;\beta) = \int \tilde{q}_{t+1}(s;\beta)\tilde{l}_{t+1}(o|s;\beta)ds \tag{15}$$

The candidate nullification strength to be applied at the next time step is then defined to minimize the surprise of the current observation $o_t$ under this candidate predictive distribution:

$$\tilde{\beta}_{t+1} = \arg\min_{0 \le \beta \le 1} \left[ -\log \tilde{p}_{t+1}\left(o_t;\beta\right) \right] \tag{16}$$

This criterion evaluates how consistently the predictive structure as a whole can accommodate current observations after simultaneously relaxing the belief and likelihood constraints. When observations could be adequately explained by the current predictive structure, a small nullification strength was selected. By contrast, when a prediction error is difficult to explain within the current predictive structure, a larger nullification strength is selected. The specific conditions under which a positive nullification strength is selected for the Gaussian model are derived as follows.

## 2.2 BIB model under a one-dimensional Gaussian assumption

The formulation introduced in the previous section is specialized for one-dimensional Gaussian distributions. For the L-step, we adopt a restricted family in which only the variance is updated, whereas the center of the likelihood is held fixed to preserve the direct correspondence between the observation and the latent state.

### 2.2.1 Distributional assumptions and the B-step

Let the belief distribution and likelihood at time $t$ be given by

$$\begin{aligned} q_t\left(s\right) &= \mathcal{N}\left(s;m_t,P_t\right), \\ l_t\left(o|s\right) &= \mathcal{N}\left(o;s,R_t\right). \end{aligned} \tag{17}$$

Here, $m_t$, $P_t$, and $R_t$ denote the belief mean, belief variance, and likelihood variance, respectively. The predictive distribution for the observation is then $p_t\left(o\right) = \int q_t\left(s\right) l_t\left(o|s\right) ds = \mathcal{N}\left(o;m_t,P_t+R_t\right)$.

In the Gaussian setting considered below, $u(s)$ is treated as a flat reference measure and its contribution is absorbed into the normalization constant. Therefore, the candidate belief generated by the B-step is $\tilde{q}_{t+1}(s;\beta) \propto q_t(s)^{1-\beta} l_t(o_t|s)$, which is again Gaussian.

The post-update belief variance is

$$\tilde{P}_{t+1}(\beta) = \frac{P_t R_t}{P_t + (1-\beta)R_t}. \tag{18}$$

Defining the learning rate as

$$K_t(\beta) = \frac{P_t}{P_t + (1-\beta)R_t}, \tag{19}$$

and the prediction error as

$$\delta_t = o_t - m_t, \tag{20}$$

the post-update belief mean is

$$\tilde{m}_{t+1}(\beta) = m_t + K_t(\beta)\delta_t. \tag{21}$$

When $\beta = 0$,

$$K_t(0) = \frac{P_t}{P_t + R_t}, \tag{22}$$

which coincides with the standard Bayesian update. For $\beta > 0$, the constraint imposed by the prior belief is weakened, resulting in an increased learning rate.

### 2.2.2 L-step

Substituting Gaussian distributions into the general L-step update rule causes the center of the likelihood to shift. In the present study, however, to preserve the direct correspondence $o \simeq s$ between the observation and the latent state, we restricted the candidate likelihood to the family

$$\tilde{l}_{t+1}\left(o \mid s;\beta\right)=\mathcal{N}\left(o;s,\tilde{R}_{t+1}\left(\beta\right)\right). \tag{23}$$

Minimizing $F_L[l;\beta]$ with respect to the likelihood variance within this restricted family yields

$$\frac{1}{\tilde{R}_{t+1}\left(\beta\right)}=\frac{1}{R_t}-\frac{\beta}{P_t+R_t}. \tag{24}$$

Therefore,

$$\tilde{R}_{t+1}\left(\beta\right)=\frac{R_t\left(P_t+R_t\right)}{P_t+\left(1-\beta\right)R_t}. \tag{25}$$

When $\beta=0$, the likelihood variance remains unchanged. In contrast, when $\beta>0$, the likelihood variance increases, thereby reducing the certainty of the correspondence between the observation and the latent state. Thus, within this restricted Gaussian family, the flattening of the predictive mass induced by the general L-step is realized as an expansion of the likelihood variance, which weakens the information provided by the observation of the latent state.

### 2.2.3 Simultaneous variance expansion and preservation of the variance ratio

In the B-step, the belief variance increases relative to the standard Bayesian update with $\beta=0$, thereby increasing responsiveness to the observation. By contrast, in the L-step, the likelihood variance increases, thereby weakening the information carried by the observation.

Taking the ratio of the candidate post-update likelihood variance to the candidate post-update belief variance yields

$$\frac{\tilde{R}_{t+1}(\beta)}{\tilde{P}_{t+1}(\beta)} = \frac{P_t + R_t}{P_t}, \tag{26}$$

which is independent of $\beta$. When $\beta = 0$, $\tilde{P}_{t+1}(0) = \frac{P_t R_t}{P_t + R_t}$ and $\tilde{R}_{t+1}(0) = R_t$.

The common expansion factor relative to the variances obtained under the standard Bayesian update is defined as $\rho_t(\beta) = \frac{P_t + R_t}{P_t + (1-\beta) R_t}$, we have $\tilde{P}_{t+1}(\beta) = \rho_t(\beta)\tilde{P}_{t+1}(0)$ and $\tilde{R}_{t+1}(\beta) = \rho_t(\beta)\tilde{R}_{t+1}(0)$.

Therefore,

$$\frac{\tilde{P}_{t+1}(\beta)}{\tilde{P}_{t+1}(0)} = \frac{\tilde{R}_{t+1}(\beta)}{\tilde{R}_{t+1}(0)} = \rho_t(\beta). \tag{27}$$

Thus, in the Gaussian setting, nullification expands the belief variance and likelihood variance obtained under the standard Bayesian update with $\beta = 0$ by the same factor. Consequently, introducing $\beta$ does not alter the ratio of the two post-update variances.

### 2.2.4 Closed-form solution for the nullification strength

Under the assumption of one-dimensional Gaussian distributions, the nullification strength can be obtained in a closed form by minimizing the predictive surprise after the candidate update. The candidate post-update predictive distribution is $\tilde{P}_{t+1}(o;\beta) = \mathcal{N}\left(o; \tilde{m}_{t+1}(\beta), \tilde{P}_{t+1}(\beta) + \tilde{R}_{t+1}(\beta)\right)$.

The candidate post-update predictive surprise for the current observation $o_t$ is

$$-\log \tilde{p}_{t+1}(o_t;\beta) = \frac{1}{2}\log\left[\frac{2\pi R_t(2P_t+R_t)}{P_t+(1-\beta)R_t}\right] + \frac{\delta_t^2(1-\beta)^2 R_t}{2(2P_t+R_t)\{P_t+(1-\beta)R_t\}}. \quad (28)$$

The candidate nullification strength is defined as $\tilde{\beta}_{t+1} = \arg\min_{0\le\beta\le1}\left[-\log \tilde{p}_{t+1}(o_t;\beta)\right]$.

Letting $x = 1-\beta$ and solving $\frac{\partial}{\partial x}\log \tilde{p}_{t+1}(o_t;x) = 0$ gives $(P_t + xR_t)(2P_t + R_t) = \delta_t^2 x(2P_t + xR_t)$. The solution is

$$x^* = \frac{R_t(2P_t+R_t) - 2P_t\delta_t^2 + \sqrt{R_t^2(2P_t+R_t)^2 + 4P_t^2\delta_t^4}}{2R_t\delta_t^2}. \quad (29)$$

Restricting the nullification strength to the interval $[0,1]$, the candidate nullification strength is given by

$$\tilde{\beta}_{t+1} = \begin{cases} 0, & \delta_t^2 \le P_t + R_t, \\ 1-x^*, & \delta_t^2 > P_t + R_t. \end{cases} \quad (30)$$

Thus, nullification is triggered not by the absolute magnitude of the prediction error but when the prediction error is large relative to the current predictive uncertainty.

## 2.3 Delayed application of the nullification candidate and the reset mechanism

It is difficult to determine whether a single large prediction error reflects a persistent change in the latent state or is merely a transient outlier. Therefore, in the present study, the candidate nullification strength computed from the current observation is not immediately applied to the update for that same observation but is instead retained as a candidate for the next time step. This corresponds to entering a state of vigilance, in which the attribution of the prediction error remains unresolved until a subsequent observation becomes available.

In the Gaussian setting considered here, nullification simultaneously increases the belief and likelihood variances, thereby increasing the uncertainty in the predictive structure as a whole. In particular, if the increase in the likelihood variance persists excessively, the correspondence between observations and latent states becomes less certain, potentially reducing the ability to infer the latent state from new observations.

Accordingly, once the next observation becomes available, the model compares the predictive surprise resulting from two alternative updates: an update that adopts the nullification candidate retained from the preceding time step, and a reset update that discards that candidate and returns to standard Bayesian updating. If standard Bayesian updating is selected, the retained nullification candidate is discarded, and the state of vigilance is terminated.

Specifically, when observation $o_t$ is obtained at time $t$, the following two hypothetical updates are compared with respect to the candidate $(\tilde{R}_t, \tilde{\beta}_t)$ retained from the preceding time step:

- a nullification update that adopts the retained candidate $(\tilde{R}_t, \tilde{\beta}_t)$;
- a reset update that discards the retained candidate and returns to $(R_0, 0)$.

Here, $R_0$ denotes the baseline value of the likelihood variance. For each condition, a post-update predictive distribution was constructed and the predictive surprise for the current observation $o_t$ is evaluated. If the predictive surprise under the reset update is smaller than that under the nullification update, the reset update is adopted.

$$\left(R_t,\beta_t\right)=\begin{cases}\left(R_0,0\right), & -\log\tilde{p}_{t+1}\left(o_t;R_0,0\right)\le-\log\tilde{p}_{t+1}\left(o_t;\tilde{R}_t,\tilde{\beta}_t\right),\\ \left(\tilde{R}_t,\tilde{\beta}_t\right), & -\log\tilde{p}_{t+1}\left(o_t;R_0,0\right)>-\log\tilde{p}_{t+1}\left(o_t;\tilde{R}_t,\tilde{\beta}_t\right).\end{cases} \tag{31}$$

The reset considered in this study does not return the belief distribution itself to its initial state. Rather, it discards the nullification candidate retained from the preceding time step and restores the nullification strength to $\beta_t = 0$ and the likelihood variance to its baseline value $R_0$, thereby returning to standard Bayesian updating from the current belief state.

Once $(R_t, \beta_t)$ to be used at time $t$ have been determined by the above criterion, the belief is updated for the current observation $o_t$ using these values. At the same time, based on the pre-update belief and the current observation $o_t$, a candidate nullification strength $\tilde{\beta}_{t+1}$ for the next time step is computed by minimizing predictive surprise as described in the preceding section, and is retained together with the corresponding candidate likelihood variance $\tilde{R}_{t+1}$. Therefore, the nullification candidate generated by the observation at time $t$ does not retrospectively alter the update at time $t$. Instead, its adoption or rejection is evaluated only when the observation at time $t+1$ becomes available.

Following a changepoint, subsequent observations are generated from the post-change state, making the retained nullification candidate more likely to be adopted. In this case, the relaxation of the belief constraint increases the learning rate and promotes the tracking of the changed state.

In contrast, following an outlier, the subsequent observation tends to return to the range corresponding to the pre-outlier state, making the reset update more likely to be selected. Consequently, the nullification candidate formed in response to the outlier is discarded, preventing its influence from persisting in subsequent time steps.

In this way, BIB does not immediately commit to the attribution of a large prediction error when it occurs but instead retains it in the form of a nullification candidate. Through the subsequent adoption or rejection of this candidate based on later observations, the response becomes temporally differentiated, after which the model shifts toward an adaptive response that facilitates tracking. In contrast, after an outlier, it shifts toward a response that suppresses the persistence of the outlier's influence.

# 3 Simulation

## 3.1 Task

To evaluate the estimation performance of the inference agents, we used a mean estimation task in a non-stationary environment containing abrupt changes in the latent mean and transient outliers, in addition to ordinary observation noise.

At each time step $t$, exactly one of three mutually exclusive events occurred: a changepoint, an outlier, or an ordinary observation. A changepoint occurred with probability $H^*$, an outlier with probability $p_O^*$, and an ordinary observation with probability $1 - H^* - p_O^*$.

When a changepoint occurred, the observation $o_t$ was sampled from a uniform distribution over an interval $[0,100]$, and this observation became the new latent mean. That is, $o_t \sim \mathcal{U}(0,100)$, $\mu_t = o_t$.

When an outlier occurred, the observation was sampled from a uniform distribution over $[0,100]$. However, the latent mean did not change and retained its value from the preceding time point: $o_t \sim \mathcal{U}(0,100)$, $\mu_t = \mu_{t-1}$

For an ordinary observation, the latent mean retained its preceding value, and the observation was generated from a Gaussian distribution centered on that latent mean: $\mu_t = \mu_{t-1}$, $o_t \sim \mathcal{N}(\mu_{t-1}, \sigma^2)$

Accordingly, the generative process for the latent mean and observation is given by

$$(\mu_t, o_t) = \begin{cases} (o_t, o_t), & o_t \sim \mathcal{U}(0,100) \quad \text{with probability } H^*, \\ (\mu_{t-1}, o_t), & o_t \sim \mathcal{U}(0,100) \quad \text{with probability } p_o^*, \\ (\mu_{t-1}, o_t), & o_t \sim \mathcal{N}(\mu_{t-1}, \sigma^2) \quad \text{with probability} 1 - H^* - p_o^*. \end{cases} \tag{32}$$

The sets of time points at which changepoints and outliers occur are denoted as $T_{CP}$ and $T_{OL}$, respectively.

In this study, three changepoint probabilities were considered: $H^* \in \{0.001, 0.01, 0.05\}$.

The outlier probability was set to $p_O^* = H^*$ and the observation-noise variance for ordinary observations was set to $\sigma^2 = 100$.

Under this setting, observations at the changepoints and outliers were generated from the same uniform distribution $\mathcal{U}(0,100)$. Thus, the likelihood of a single observation is identical for the two event types, and the observation value itself contains no information that distinguishes a changepoint from an outlier.

The difference between the two event types lies in the persistence of the latent mean after an event. In the case of a changepoint, the observation $o_t$ at the event becomes the new latent mean, and subsequent ordinary observations are generated around this new mean. In contrast, in the case of an outlier, the latent mean remains unchanged, and ordinary observations from the next time point return to the neighborhood of the original mean.

## 3.2 Inference agents

Five types of inference agents were compared using the mean estimation task described in the preceding section.

The first is a BIB agent that implements the proposed model.

The second is the SH agent, which is based on the Sage–Husa-type adaptive Kalman filter, in which the process-noise variance is sequentially adapted (Sage and Husa 1969). In this study, the observation-noise variance was fixed at the true value used in the generative process and only the process-noise variance was adapted.

The third is the RB agent based on the reduced Bayesian model proposed by Nassar et al. (2010). Drawing on the normative framework of Nassar et al. (2019), in which changepoints and transient outliers (oddballs) were treated as distinct statistical contexts, we extended the model to consider three hypotheses within a single sequence: ordinary observation, changepoint, and outlier.

The fourth is the fixed-$\beta$ BIB agent, in which only the computation of the candidate nullification strength in BIB is replaced by an externally specified constant, $\beta_0$. The delayed application and reset criterion are identical to those of the BIB agent. Therefore, a comparison with the BIB agent allows us to evaluate the effect of endogenously determining the nullification strength.

The fifth is the forgetting Bayesian (FB) agent, in which the likelihood variance is fixed at its baseline value $R_0$, whereas forgetting on the belief side is applied at every time step. Comparison with the fixed-$\beta$ BIB agent allows us to evaluate the effect of the control mechanism consisting of likelihood-variance expansion and recovery through resetting.

The details of each agent are described below.

### 3.2.1 BIB agent

One step of the BIB agent consists of a reset decision followed by the BIB update. In the reset decision, the agent determines whether to retain the nullification candidate carried over from the previous time step or to reset to standard Bayesian updating according to the predictive-surprise criterion given in Eq. (31).

Using the values selected by this criterion, the belief mean, belief variance, candidate likelihood variance, candidate nullification strength, and learning rate are updated as follows:

$$
\begin{aligned}
m_{t+1} &= m_t + K_t\delta_t, \\
P_{t+1} &= \frac{P_t R_t}{P_t + (1-\beta_t) R_t}, \\
\tilde{R}_{t+1} &= \frac{R_t (P_t + R_t)}{P_t + (1-\beta_t) R_t}, \\
\tilde{\beta}_{t+1} &= \begin{cases} 0, & \delta_t^2 \le P_t + R_t, \\ 1 - \dfrac{R_t(2P_t + R_t) - 2P_t\delta_t^2 + \sqrt{R_t^2(2P_t + R_t)^2 + 4P_t^2\delta_t^4}}{2R_t\delta_t^2} & \delta_t^2 > P_t + R_t, \end{cases} \\
K_t &= \frac{P_t}{P_t + (1-\beta_t) R_t}.
\end{aligned} \tag{33}
$$

Here, $\delta_t = o_t - m_t$ is the prediction error.

### 3.2.2 SH agent

The SH agent assumes a one-dimensional random-walk model, with the predicted variance given by $P_t^-$. The learning rate (Kalman gain) is $K_t$. The state estimate $m_{t+1}$, estimation-error variance $P_{t+1}$, the process-noise variance $Q_{t+1}$, and observation-noise variance $R_{t+1}$ are updated according to

$$
\begin{aligned}
P_t^- &= P_t + Q_t, \\
K_t &= \frac{P_t^-}{P_t^- + R_t}, \\
m_{t+1} &= m_t + K_t\delta_t, \\
P_{t+1} &= (1-K_t) P_t^-, \\
Q_{t+1} &= \max\left\{0, (1-\alpha_Q) Q_t + \alpha_Q \left(K_t^2\delta_t^2 + P_{t+1} - P_t\right)\right\}, \\
R_{t+1} &= \max\left\{0, (1-\alpha_R) R_t + \alpha_R \left(\delta_t^2 - P_t^-\right)\right\}.
\end{aligned} \tag{34}
$$

Here, $\alpha_Q$ and $\alpha_R$ denote the adaptation rates for the process-noise variance and observation-noise variance, respectively.

### 3.2.3 RB agent

The RB agent assumes that observation $o_t$ at each time step is generated by one of three mutually exclusive events: nominal observation, changepoint, or outlier. A changepoint occurs with probability $H$, an outlier with probability $p_o$, and a nominal observation with probability $1 - H - p_o$.

The RB agent is given the true values of changepoint probability $H$, outlier probability $p_o$, and observation-noise variance $R_0$ for nominal observations.

Under the nominal-observation hypothesis, the current belief about the latent mean is $q_t(\mu) = \mathcal{N}(\mu; m_t, P_t)$, and the observation model is $o_t \mid \mu_t \sim \mathcal{N}(\mu_t, R_0)$.

The predictive likelihood under the nominal-observation hypothesis is therefore $L_t = \frac{1}{\sqrt{2\pi(P_t+R_0)}} \exp\left[-\frac{\delta_t^2}{2(P_t+R_0)}\right]$, where $\delta_t = o_t - m_t$ is the prediction error. The gain under the nominal-observation hypothesis is $\alpha_t = \frac{P_t}{P_t+R_0}$.

Under both the changepoint and outlier hypotheses, the observation is assumed to be generated from a uniform distribution over an interval $[0,100]$. Thus, the observation likelihood $U(o_t)$ is the same under these two hypotheses. Let $\Omega_t$, $O_t$, and $\Gamma_t$ denote the posterior probabilities of the changepoint, outlier, and nominal-observation hypotheses, respectively. They are given by $\Omega_t = \frac{HU(o_t)}{Z_t}$, $O_t = \frac{p_o U(o_t)}{Z_t}$, $\Gamma_t = \frac{(1-H-p_o)L_t}{Z_t}$, where the normalization constant is given by $Z_t$.

Because the observation likelihoods are identical under the changepoint and outlier hypotheses, the ratio of their posterior probabilities is $\frac{\Omega_t}{O_t} = \frac{H}{p_o}$. In particular, when $H = p_o$ as in the present study, $\Omega_t = O_t$ always holds. Thus, the RB agent cannot distinguish a changepoint from an outlier based solely on observations at the time of the event.

Under the changepoint hypothesis, the observation $o_t$ is regarded as the new latent mean. Accordingly, the posterior mean and variance under the changepoint hypothesis are set to $m_{t+1}^{CP} = o_t$ and $P_{t+1}^{CP} = 0$. Under the nominal-observation hypothesis, the standard Kalman update gives $m_{t+1}^{N} = m_t + \alpha_t \delta_t$ and $P_{t+1}^{N} = (1 - \alpha_t)P_t$. By contrast, under the outlier hypothesis, the observation is not used to estimate the latent mean, and the pre-update belief is retained. Thus, $m_{t+1}^{O} = m_t$ and $P_{t+1}^{O} = P_t$. The mean estimate obtained by weighting the three hypothesis-specific posterior means by their posterior probabilities is $m_{t+1} = \Omega_t o_t + \Gamma_t(m_t + \alpha_t \delta_t) + O_t m_t$. Defining the effective learning rate as $K_t = \Omega_t + \Gamma_t \alpha_t$ , the resulting update equations are given by

$$
\begin{aligned}
&m_{t+1} = m_t + K_t \delta_t. \\
&K_t = \Omega_t + \Gamma_t \alpha_t, \\
&\alpha_t = \frac{P_t}{P_t + R_0}, \\
&\Omega_t = \frac{HU(o_t)}{Z_t}, \\
&O_t = \frac{p_o U(o_t)}{Z_t}, \\
&\Gamma_t = \frac{(1 - H - p_o)L_t}{Z_t}, \\
&Z_t = (1 - H - p_o)L_t + (H + p_o)U(o_t), \\
&L_t = \frac{1}{\sqrt{2\pi(P_t + R_0)}} \exp\left[-\frac{\delta_t^2}{2(P_t + R_0)}\right], \\
&U(o_t) = \begin{cases} \frac{1}{100}, & 0 \le o_t \le 100, \\ 0, & otherwise, \end{cases} \\
&P_{t+1} = \Gamma_t(1 - \alpha_t)P_t + O_t P_t + \left[\Omega_t(1 - K_t)^2 + \Gamma_t(\alpha_t - K_t)^2 + O_t K_t^2\right]\delta_t^2.
\end{aligned}
\tag{35}
$$

The posterior variance $P_{t+1}$ is obtained from the law of total variance applied to the mixture distribution for the following three hypotheses. The first term on the right-hand side corresponds to the within-hypothesis variance under the nominal observation hypothesis, whereas the second term corresponds to the retained prior variance under the outlier hypothesis. Under the changepoint hypothesis, the within-hypothesis variance is zero because the latent mean is uniquely determined by $o_t$. The third term represents the between-hypothesis variance arising from the differences between the hypothesis-specific posterior means and mixture mean $m_{t+1}$.

### 3.2.4 Fixed-β BIB agent

The fixed-$\beta$ BIB agent is a variant of the BIB agent in which only the procedure for determining the candidate nullification strength is replaced by an exogenously specified constant. In the BIB agent, the candidate nullification strength for the next time step, $\tilde{\beta}_{t+1}$, is determined in closed form from the current prediction error $\delta_t$ and the pre-update variances $P_t$ and $R_t$. In contrast, in the fixed-$\beta$ BIB agent, $\tilde{\beta}_{t+1} = \beta_0$, where $\beta_0 \in [0,1]$ is a constant independent of both the prediction error and the variances.

The update rule for the candidate likelihood variance, delayed application of the nullification candidate, and reset decision are identical to those of the BIB agent. Thus, in the fixed-$\beta$ BIB agent, nullification is applied only at time steps at which the continuation candidate is selected; when the reset candidate is selected, standard Bayesian updating is performed with $\beta_t = 0$.

A comparison with a BIB agent allows us to evaluate the effect of the mechanism that endogenously determines the candidate nullification strength according to the current prediction error and uncertainty. The two agents differ only in the rule used to generate the candidate nullification strength; whether the candidate is applied at the next time step is determined by the post-update predictive surprise for the subsequent observation.

### 3.2.5 FB agent

The FB agent fixes the likelihood variance at its baseline value and applies forgetting only to belief at every time step. Specifically, for all the time steps, $R_t = R_0$, $\beta_t = \beta_0$ and the belief mean and variance are updated according to

$$\begin{aligned} m_{t+1} &= m_t + K_t \delta_t, \\ K_t &= \frac{P_t}{P_t + (1-\beta_0) R_0}, \\ P_{t+1} &= \frac{P_t R_0}{P_t + (1-\beta_0) R_0} \end{aligned} \tag{36}$$

Thus, the FB agent corresponds to Bayesian inference with a fixed forgetting strength $\beta_0$.

Under this update rule, the belief variance has the fixed point $P^* = \beta_0 R_0$ and the corresponding learning rate is $K^* = \frac{\beta_0 R_0}{\beta_0 R_0 + (1-\beta_0) R_0} = \beta_0$.

Thus, in the FB agent, the steady-state learning rate is exactly equal to the forgetting strength $\beta_0$.

Comparison with the fixed-$\beta$ BIB agent allows us to evaluate the effect of the control mechanism consisting of likelihood-variance expansion and recovery through resetting.

### 3.2.6 Initial conditions and parameter settings

The following initial values were used for all agents:

$$\begin{aligned} &\mu_0 \sim \mathcal{U}(0,100), \\ &m_0 \sim \mathcal{U}(0,100), \\ &P_0 = 1000, \\ &R_0 = \sigma^2 = 100. \end{aligned} \tag{37}$$

Thus, the initial observation-noise variance (likelihood variance) $R_0$ was set as the true value used in the generative process. In each trial, all agents were presented with the same latent means and observation sequences.

The SH agent has two adaptation parameters, $\alpha_R$ and $\alpha_Q$. Because the true observation-noise variance was provided to the agent, $\alpha_R$ was set to $0$. Under this setting, the SH agent did not update its estimate of the observation-noise variance, which therefore remained fixed at the true value $R_0 = \sigma^2$. In contrast, the adaptation rate for the process-noise variance, $\alpha_Q$, was varied over $\alpha_Q \in [0,1]$ in increments of $0.02$. The initial process-noise variance was set to $Q_0 = 0$.

The fixed-β BIB and FB agents share a fixed strength parameter $\beta_0$, interpreted as the nullification strength in the fixed-β BIB agent and as the forgetting strength in the FB agent. This parameter was varied over $\beta_0 \in [0,1]$ in increments of $0.02$.

For the RB agent, the changepoint and outlier probabilities were set to their true generative-process values: $H = H^*, \quad p_o = p_o^*$.

The uniform likelihood under the changepoint and outlier hypotheses was defined over the same observation range $[0,100]$ as in the generative process, giving $U(o_t) = \frac{1}{100}$.

Thus, the RB agent was evaluated under an oracle setting, in which the major parameters of the generative process were assumed to be known. Importantly, this oracle setting means that the generative parameters are known and not that the event type occurring at each time step is directly given to the agent.

Each simulation consisted of 105,000 timesteps. To exclude the influence of the initial transient, the first 5,000 time steps were discarded, leaving 100,000 time steps per trial for the analysis. The simulation was repeated for 1000 trials using different random seeds.

The simulations were implemented in C++ using MinGW 11.2.0 (64-bit) and the Qt 6.6.1 library. Random numbers were generated using the Mersenne Twister pseudorandom-number generator (std::mt19937). Uniform random variates were generated using std::uniform_real_distribution, and Gaussian random variates were generated from the same pseudorandom-number generator. To ensure reproducibility across simulation trials, the 1000 trials were initialized using distinct integer seeds from 1 to 1000, with one seed assigned to each trial.

### 3.2.7 Evaluation metrics

The evaluation was conducted from three perspectives. For all the agents, $m_{t+1}$ denotes the belief mean after incorporating observation $o_t$. Therefore, we define $\tilde{\mu}_t \equiv m_{t+1}$ as the estimate of the latent mean at time $t$ based on observations up to and including $o_t$.

The first analysis examined the learning-rate response along an event-aligned time axis, following changepoints and outliers. Let $t_e$ denote the time at which an event occurs and let $\tau = t - t_e$ denote the time relative to the event, such that $\tau = 0$ corresponds to the event time. For each $\tau = 0,1,\ldots,100$, the learning rate was first averaged across all events within each trial and then averaged across trials. The resulting value was defined as the learning-rate time profile $\bar{K}(\tau)$.

If another event, either a changepoint or an outlier, occurred within $\tau \leq 100$, the corresponding profile was truncated at that point and did not contribute to subsequent values of $\tau$. Consequently, the number of samples that contributed to the average decreased as $\tau$ increased. From the resulting post-changepoint profile $\bar{K}_{CP}(\tau)$ and post-outlier profile $\bar{K}_{OL}(\tau)$, we computed the difference $\Delta\bar{K}(\tau)$ at each relative time to evaluate how each agent differentiated its internal response to changepoint and outlier events.

$$\begin{aligned}
&\bar{K}_{CP}(\tau) = E\left[K_{t_e+\tau} | t_e \in T_{CP}\right], \\
&\bar{K}_{OL}(\tau) = E\left[K_{t_e+\tau} | t_e \in T_{OL}\right], \\
&\Delta\bar{K}(\tau) = \bar{K}_{CP}(\tau) - \bar{K}_{OL}(\tau).
\end{aligned} \tag{38}$$

In addition to the learning rate, we computed event-aligned profiles of three internal variables of the BIB agent: the applied nullification strength $\bar{\beta}(\tau)$, the reset selection rate $\bar{r}(\tau)$, and the applied likelihood variance normalized by its baseline value, $\bar{R}(\tau)/R_0$. Here $\beta_t$ and $R_t$ denote the values adopted after the reset decision, not the candidates ($\tilde{\beta}_{t+1}$, $\tilde{R}_{t+1}$) generated at that step. The reset selection rate is defined as the proportion of events for which, at relative time $\tau$, the belief was updated with $(R_0, 0)$. Therefore, it counts not only the time steps at which a retained nullification candidate was explicitly rejected in favor of the reset candidate but also the time steps at which the retained candidate was itself $(R_0, 0)$, so that the update coincided with standard Bayesian updating at the baseline likelihood variance. Note that the two update candidates are strictly identical only when $\tilde{\beta}_t = 0$ and $\tilde{R}_t = R_0$ hold simultaneously; when $\tilde{\beta}_t = 0$ but $\tilde{R}_t > R_0$, the candidates still differ in their post-update predictive variance, and the comparison of predictive surprise remains effective. Averaging within and across trials and truncation at the next event followed the same rules as those used for learning rate profiles.

The second analysis evaluated the estimation performance using the event-specific mean squared error (MSE). At each relative time $\tau$, squared errors were first averaged across all events within each trial and then averaged across trials. This yielded the post-changepoint MSE time profile $MSE_{CP}(\tau)$ and the post-outlier MSE time profile $MSE_{OL}(\tau)$.

$$
\begin{aligned}
MSE_{CP}(\tau) &= E\left[\left(\tilde{\mu}_{t_e+\tau} - \mu_{t_e+\tau}\right)^2 \mid t_e \in T_{CP}\right], \\
MSE_{OL}(\tau) &= E\left[\left(\tilde{\mu}_{t_e+\tau} - \mu_{t_e+\tau}\right)^2 \mid t_e \in T_{OL}\right].
\end{aligned}
\tag{39}
$$

The same truncation rule used for the learning rate profiles was applied. These MSE time profiles directly characterize the temporal evolution of the estimation error following each type of event.

We then defined the cumulative MSE over $\tau = 0$ to $100$ as

$$\begin{aligned} M_{CP} &= \sum_{\tau=0}^{100} MSE_{CP}(\tau), \\ M_{OL} &= \sum_{\tau=0}^{100} MSE_{OL}(\tau). \end{aligned} \tag{40}$$

A smaller $M_{CP}$ indicates better tracking following a changepoint, whereas a smaller $M_{OL}$ indicates greater stability following an outlier. The location of each agent in the $(M_{OL}, M_{CP})$ plane was used to evaluate the trade-off between rapid error reduction after a changepoint and the suppression of error increase following an outlier.

The learning-rate profile, MSE profile, and cumulative MSE analyses described above were performed for the BIB, RB, and SH agents, whereas the internal variable profiles were computed for the BIB agent only. For the SH agent, $\alpha_Q$ was swept over its parameter range, whereas the time profiles were presented for two representative conditions, $\alpha_Q = 1.0$ and $\alpha_Q = 0.5$.

The fixed-$\beta$ BIB and FB agents were introduced as ablation models to isolate the effects of the mechanism for endogenous determination of $\beta$ and the L-step in BIB. By comparing the fixed-$\beta$ BIB and FB agents, the effect of the nullification control mechanism involving likelihood-variance updating was evaluated in the cumulative MSE plane.

The third analysis evaluated the overall performance using root mean square error (RMSE) over the entire simulation period. For each trial, the squared error was averaged across all time points and these values were averaged across trials before taking the square root.

$$RMSE_{all} = \sqrt{\frac{1}{N_{trial}} \sum_{n=1}^{N_{trial}} \frac{1}{T} \sum_{t=1}^{T} \left( \tilde{\mu}_t^{(n)} - \mu_t^{(n)} \right)^2} \tag{41}$$

This measure was compared across all five agents: BIB, RB, SH, fixed-$\beta$ BIB, and FB.

Although the cumulative MSE and overall RMSE are both based on aggregated squared errors, they differ in the set of time points included in the evaluation and in their temporal weighting. Consequently, the rankings of the models are not identical for the two measures.

# 4 Simulation results

## 4.1 Post-event response characteristics of learning rate and estimation error

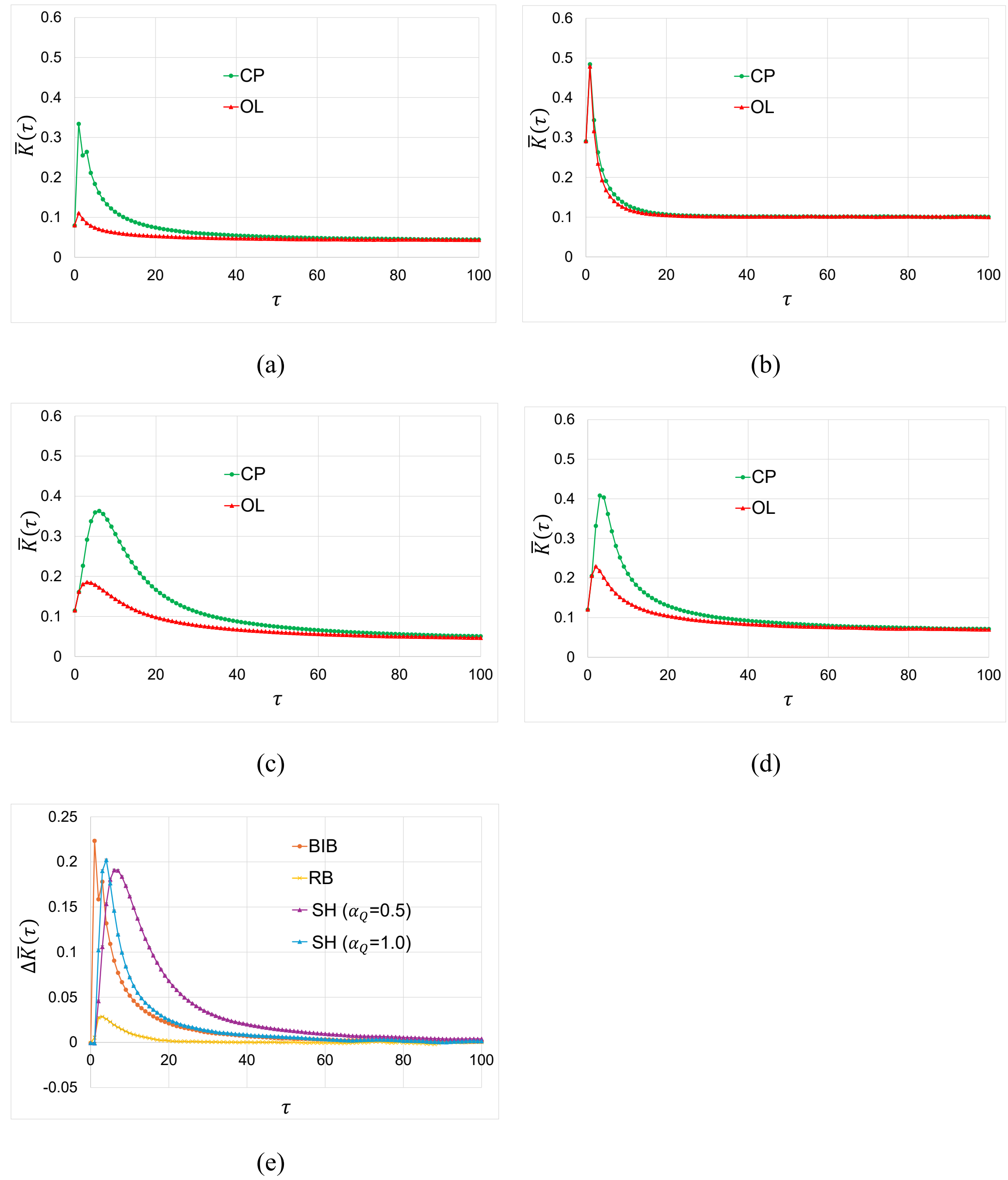

**Fig. 1** Post-event time profiles of the learning rate $K$ for the BIB, RB, and SH agents under $H^* = p_O^* = 0.01$, with SH shown at two adaptation rates. (a) BIB. (b) RB. (c) SH ($\alpha_Q = 0.5$). (d) SH ($\alpha_Q = 1.0$). (e) Difference profiles between the post-changepoint and post-outlier responses, $\Delta\bar{K}(\tau) = \bar{K}_{CP}(\tau) - \bar{K}_{OL}(\tau)$

Fig. 1 shows the learning-rate time profiles following changepoints and outliers for the four agent

configurations—BIB, RB, SH ($\alpha_Q = 0.5$), and SH ($\alpha_Q = 1.0$)—under the condition $H^* = p_o^* = 0.01$. Panels (a)–(d) of Fig. 1 overlay, for each agent, the post-changepoint profile $\overline{K}_{CP}(\tau)$ and the post-outlier profile $\overline{K}_{OL}(\tau)$, while panel (e) overlays the difference profiles for all agents, defined as $\Delta\overline{K}(\tau) = \overline{K}_{CP}(\tau) - \overline{K}_{OL}(\tau)$. The difference profile represents the extent to which each agent exhibits distinct internal responses to changepoint and outlier events.

For all agents, the learning rates following changepoints and outliers were nearly identical at $\tau = 0$. This reflects a constraint inherent in task design: because the observations at the event time were generated from the same uniform distribution, $U(0,100)$, for both event types, the two events could not be distinguished from a single observation. We therefore compare how the responses of the agents diverged at $\tau \geq 1$.

For BIB, at $\tau \geq 1$, the consistency assessment based on subsequent observations led to different responses for the two event types. Following a changepoint, the nullification candidate was more frequently retained, resulting in an increased learning rate, whereas following an outlier, the reset update was more frequently selected, suppressing the increase in learning rate. The difference profile $\Delta\overline{K}(\tau)$ showed a clear positive peak at $\tau = 1$, indicating that BIB initially suspends causal attribution and subsequently differentiates its response according to the event's consistency with subsequent observations. The dynamics of the internal variables underlying this differentiation are presented in Section 4.2.

For RB, because $H = p_o$ and the observation likelihoods under the changepoint and outlier hypotheses are also identical, $\Omega_t = O_t$ at the event time. At $\tau \geq 1$, the learning rates can, in principle, diverge depending on the subsequent observation sequence; however, in the present simulation, the difference between the post-changepoint and post-outlier profiles was small.

Similar to BIB, SH also exhibited response differentiation, but with a substantial delay in the increase in learning rate. The delay became larger as $\alpha_Q$ decreased. Even under the $\alpha_Q = 1.0$ condition, which showed the shortest delay within the scanned range, the peak of the difference profile $\Delta\bar{K}(\tau)$ occurred at $\tau = 4$, later than the BIB peak at $\tau = 1$.

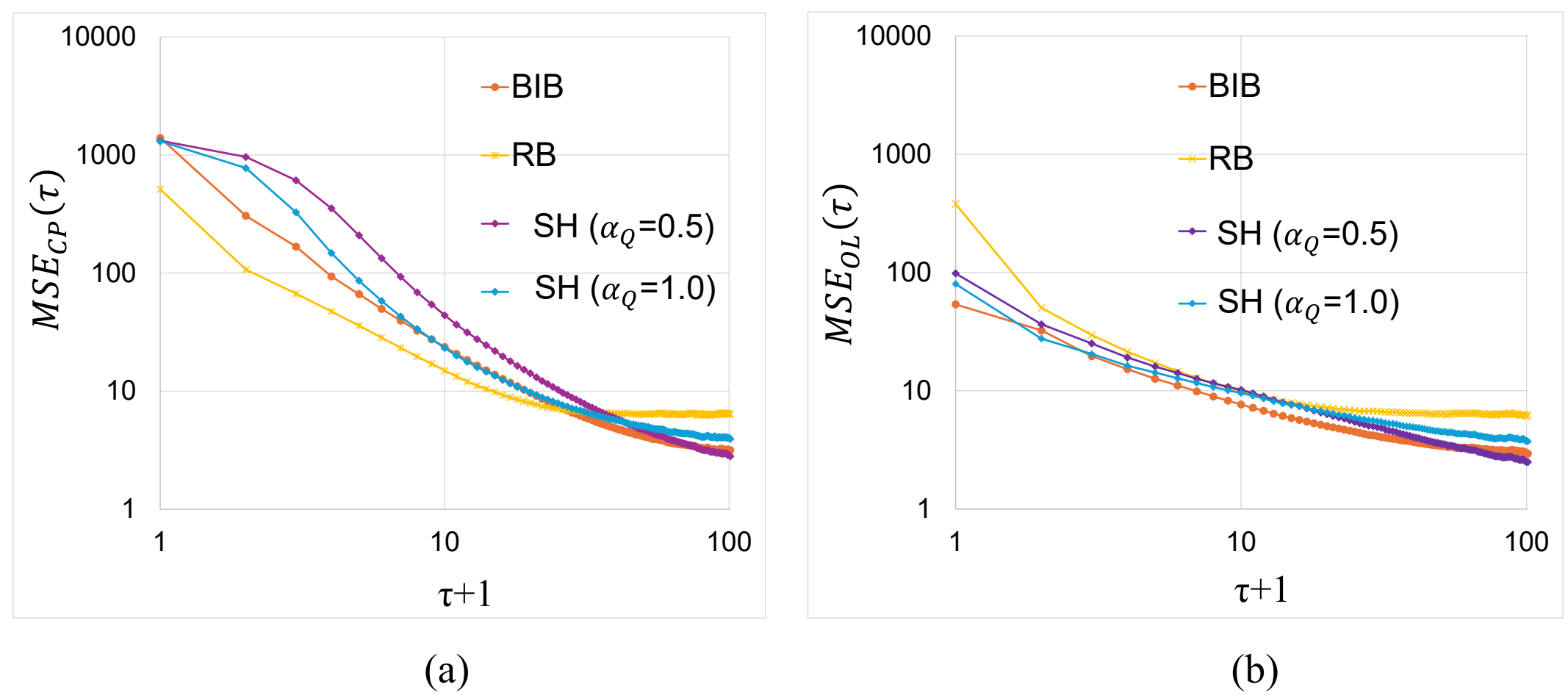

**Fig. 2** Post-event time profiles of MSE for the BIB, RB, and SH agents under $H^* = p_O^* = 0.01$. Each panel overlays the profiles of four agent configurations: BIB, RB, SH ($\alpha_Q = 0.5$), and SH ($\alpha_Q = 1.0$). Both axes are shown on logarithmic scales. To permit logarithmic scaling of the horizontal axis while retaining the event time, relative time is plotted as $\tau + 1$; thus, $\tau = 0$ is plotted at $\tau + 1 = 1$, and $\tau = 100$ at $\tau + 1 = 101$. (a) Post-changepoint MSE profile, $MSE_{CP}(\tau)$. (b) Post-outlier MSE profile, $MSE_{OL}(\tau)$

Fig. 2 shows the post-changepoint and post-outlier MSE profiles for the same four agents. Panel (a) shows the MSE profiles following changepoints, whereas panel (b) shows the MSE profiles following outliers. In each panel, the profiles of all four agents are overlaid. Whereas the learning-rate profiles represent the internal response to an event, the MSE profiles represent the actual temporal evolution of estimation error following an event.

For the post-changepoint MSE profiles (panel (a)), all agents showed a high MSE at $\tau = 0$, which decreased as $\tau$ increased. The MSE at $\tau = 0$ is determined by how closely the estimate approaches the true value ($\mu_t = o_t$) after incorporating the observation $o_t$ at the changepoint event. Agents with higher learning rates incorporate the observation more strongly at $\tau = 0$, resulting in lower MSE at that time point. Because BIB has a relatively low learning rate at $\tau = 0$, it showed the highest post-changepoint MSE at $\tau = 0$. At $\tau \geq 1$, however, the increase in learning rate resulting from the maintenance of nullification in BIB accelerated tracking of the changepoint, causing the MSE to decrease rapidly. In SH, the delayed increase in the learning rate was accompanied by a corresponding delay in the MSE reduction. RB showed the lowest MSE during the early post-change period, whereas at later lags, its MSE became slightly higher than that of BIB, SH ($\alpha_Q = 0.5$), and SH ($\alpha_Q = 1.0$). Following outliers, RB showed the highest MSE at $\tau = 0$ (panel (b)). The MSE at $\tau = 0$ reflects the extent to which the outlier observation is incorporated into the estimate. Agents with higher learning rates incorporate the outlier more strongly and therefore exhibit higher MSE at $\tau = 0$. At $\tau \geq 1$, subsequent observations are generated around the original latent mean, and the effect of the outlier is therefore gradually removed. In BIB, a reset was likely to be selected at $\tau = 1$, releasing nullification and returning the agent to standard Bayesian updating. Consequently, the influence of the outlier is rapidly suppressed. In SH, the effect of the outlier persisted for a longer period, depending on the decay time constant of $Q_t$.

## 4.2 Post-event dynamics of the internal variables of BIB

The learning-rate profiles in Fig. 1 characterize the consequences of the BIB response, rather than the internal variables that generate it. Fig. 3 shows, under the same condition $H^* = p_O^* = 0.01$, the post-event time profiles of the nullification strength, reset selection rate, and likelihood variance following changepoints and outliers.

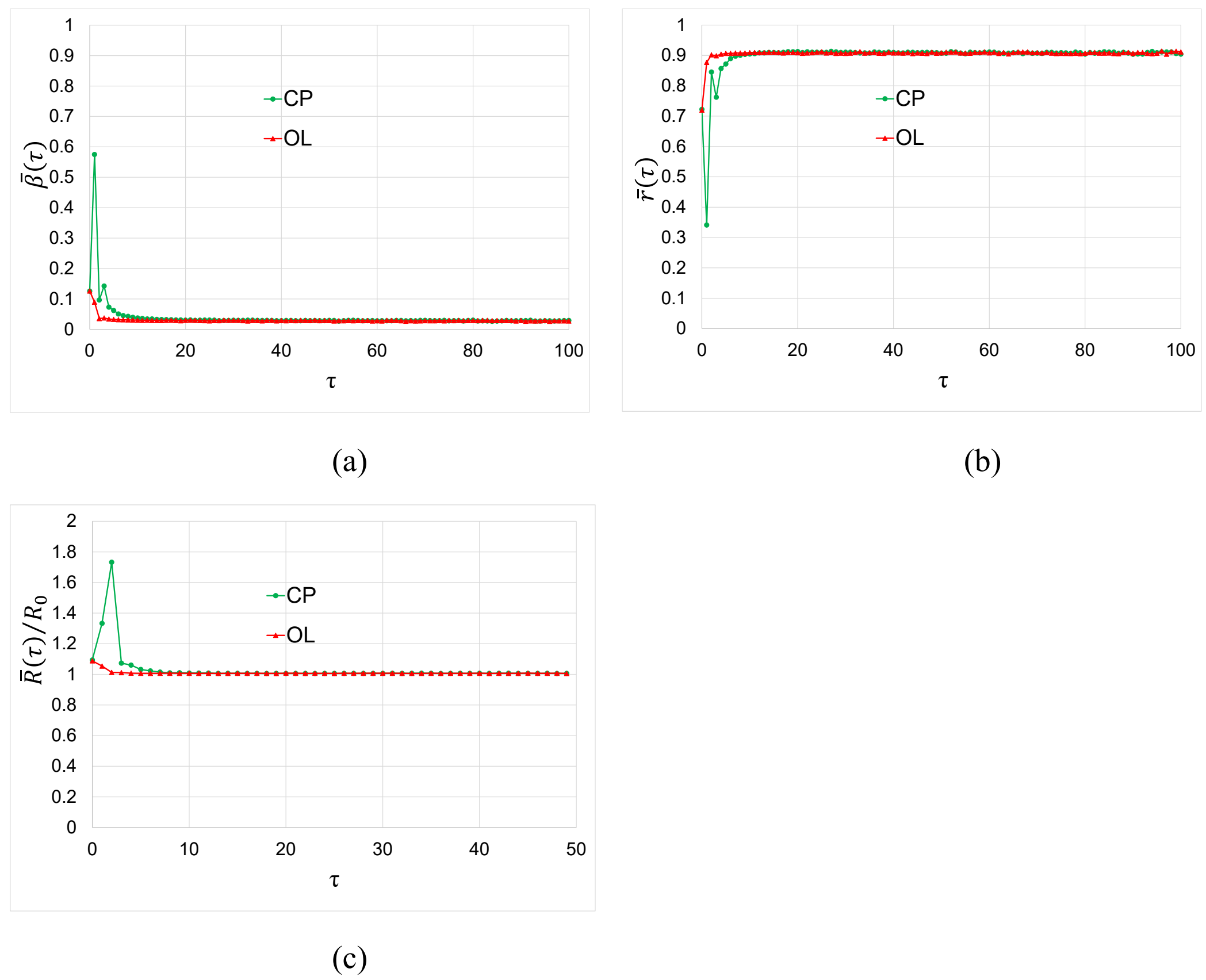

**Fig. 3** Post-event time profiles of the internal variables of the BIB agent under $H^* = p_o^* = 0.01$. In each panel, the profile following changepoints (CP) and that following outliers (OL) are overlaid. (a) Applied nullification strength, $\bar{\beta}(\tau)$. (b) Reset selection rate, $\bar{r}(\tau)$. (c) Applied likelihood variance normalized by its baseline value, $\bar{R}(\tau)/R_0$. Averaging and truncation follow the same rules as in Fig. 1

At $\tau = 0$, the applied nullification strength was the same for changepoints and outliers, with no difference between the two event types. Its value was nevertheless higher than the steady-state level: $\bar{\beta}(0) \approx 0.12$, compared with approximately 0.03 for $\tau > 10$. The reset selection rate at $\tau = 0$ was approximately 0.7, below its steady-state value of approximately 0.9, and the applied likelihood variance was correspondingly elevated, with $\bar{R}(0)/R_0 \approx 1.1$ compared with approximately 1.01 in the steady state.

This elevation did not originate from the nullification candidates generated by the prediction error of the event. Under a delayed application, the candidate is applied only at the next time step. What is applied at $\tau = 0$ is the candidate retained from the preceding ordinary observation, and the elevation therefore reflects whether that candidate is adopted or discarded. Because the reset decision compares the predictive surprise for the current observation $o_t$, a large prediction error at the event favors the nullification candidate which has a wider predictive variance. Thus, a small candidate that is ordinarily discarded is retained. Consistent with this, the retention rate increased from approximately 0.1 in the steady state to approximately 0.3 at $\tau = 0$, which largely accounts for the elevation of $\bar{\beta}$. Because the observation distribution at the event was identical for both event types, this elevation occurred equally after the changepoints and outliers.

At $\tau = 1$, the two profiles diverged sharply. The reset selection rate decreased to 0.34, and the nullification candidate generated by the large prediction error at the event was more frequently retained, increasing the mean applied nullification strength to 0.58. In contrast, following the outliers, the reset selection rate increased to 0.88, and the nullification strength decreased to 0.09. The applied likelihood variance behaved correspondingly, expanding to $\bar{R}(1)/R_0 = 1.34$ after changepoints while remaining at 1.06 after the outliers. Because $\beta_t = 0$ by definition whenever a reset is selected, $\bar{\beta}(1)$ equals the product of the retention rate and the nullification strength conditional on retention. Under this decomposition, the conditional strength was comparable across event types—approximately 0.9 after changepoints and approximately 0.8 after outliers—such that the difference between the two arises mainly from how often the candidate was retained (0.66 versus 0.12). Therefore, response differentiation reflects the adoption or rejection of a candidate rather than the magnitude of the candidates generated. This corresponds to the mechanism described in Section 2.3, because the observation following an outlier returns to the vicinity of the original latent mean, and standard Bayesian updating yields a smaller predictive surprise than an update that adopts the nullification candidate.

For $\tau \geq 2$, the nullification strength following changepoints decayed, and by $\tau > 10$ both event types converged to the steady-state level ($\bar{\beta} \approx 0.03$, reset selection rate $\approx 0.9$, $\bar{R}/R_0 \approx 1.01$). Maintaining nullification requires the continuation candidate to be selected at each subsequent step, and resets become more likely as the belief catches up with the observations generated from the post-change state. The nullification strength following outliers had already declined substantially toward the steady-state level by $\tau = 1$, although it remained approximately three times higher than the steady-state level, indicating that the influence of the outlier is released rapidly but not instantaneously. A steady-state reset selection rate of approximately 0.9 indicates that during ordinary observation periods, BIB operates essentially as a standard Bayesian updating.

These profiles show that the positive peak of the difference profile $\Delta\bar{K}(\tau)$ at $\tau = 1$ in Fig. 1(e) corresponds directly to the point at which the adoption or rejection of the nullification candidate differentiates according to event type. Although BIB does not explicitly represent changepoint and outlier hypotheses, a single mechanism—the application and release of nullification—generates event-dependent differentiation of the internal state through consistency with subsequent observations.

### 4.3 Trade-off between tracking and stability

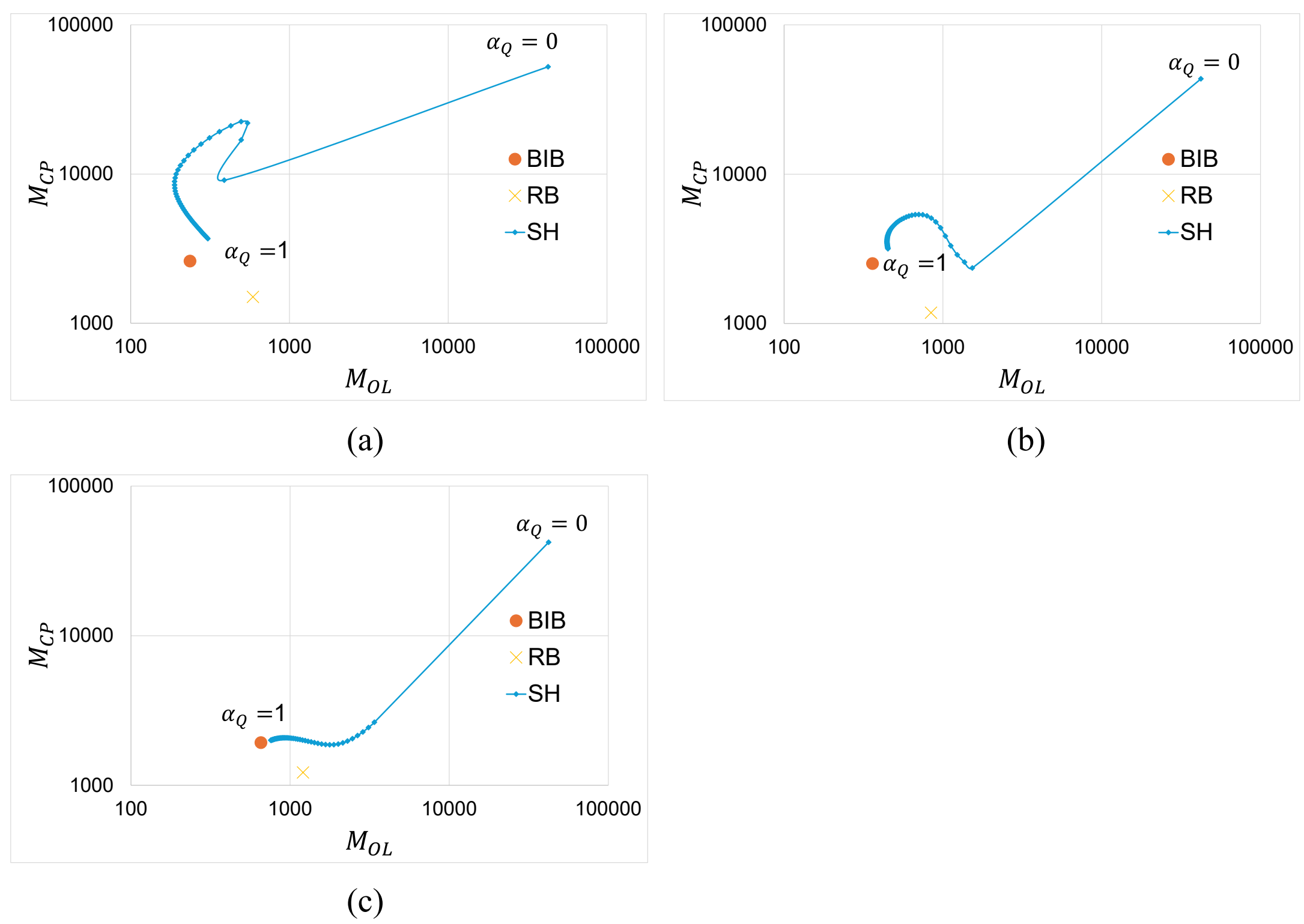

**Fig. 4** Relationship between tracking performance and stability for the BIB, RB, and SH agents. For the SH agent, the adaptation rate $\alpha_Q \in [0,1]$ was scanned in increments of 0.02. The RB agent shows the value obtained under the oracle setting. For all three agents, the initial likelihood variance was set to the true value of the generative process, $R_0 = \sigma^2 = 100$. (a) $H^* = p_o^* = 0.001$. (b) $H^* = p_o^* = 0.01$. (c) $H^* = p_o^* = 0.05$. In each panel, the vertical axis represents the cumulative post-changepoint MSE, $M_{CP}$, and the horizontal axis represents the cumulative post-outlier MSE, $M_{OL}$. Both axes are on a logarithmic scale

Fig. 4 shows the positions of BIB, RB, and SH, with $\alpha_Q$ scanned in the cumulative MSE plane $(M_{OL}, M_{CP})$ under three conditions, $H^* = p_o^* \in \{0.001, 0.01, 0.05\}$. For SH, $\alpha_Q \in [0,1]$ was scanned in increments of 0.02, and the resulting sequence of points is shown as a curve. The BIB and RB are shown as single points.

Under all three conditions, no scanned value of $\alpha_Q$ yielded both a smaller $M_{CP}$ and a smaller $M_{OL}$ than BIB; that is, BIB was nondominated with respect to the SH sweep, even though, unlike SH, it has no corresponding tuning parameter. Comparing BIB and RB, RB exhibited better tracking performance under all conditions, whereas BIB exhibited better stability.

Fig. 5 shows the positions of BIB, fixed-$\beta$ BIB with a scan over $\beta_0$, and FB with a scan over $\beta_0$ in the cumulative MSE plane under the same three conditions. The two sweeps largely coincided near their most favorable regions, and neither was uniformly closer to the origin: along the small-$\beta_0$ branch, the FB sweep lay slightly below that of fixed-$\beta$ BIB, whereas along the large-$\beta_0$ branch, the ordering was reversed, with the fixed-$\beta$ BIB sweep terminating at $\beta_0 = 1$ without extending to the large $M_{OL}$ reached by FB. Under all three conditions, no scanned value of $\beta_0$, for either agent, yielded both a smaller $M_{CP}$ and a smaller $M_{OL}$ than BIB, so the BIB point was nondominated with respect to both sweeps; under $H^* = p_o^* = 0.05$, BIB additionally dominated the entire FB sweep.

These results are consistent with the contributions of both the nullification-control mechanism, including likelihood-variance updating and resetting, and the endogenous determination of the nullification strength to the estimation performance of BIB.

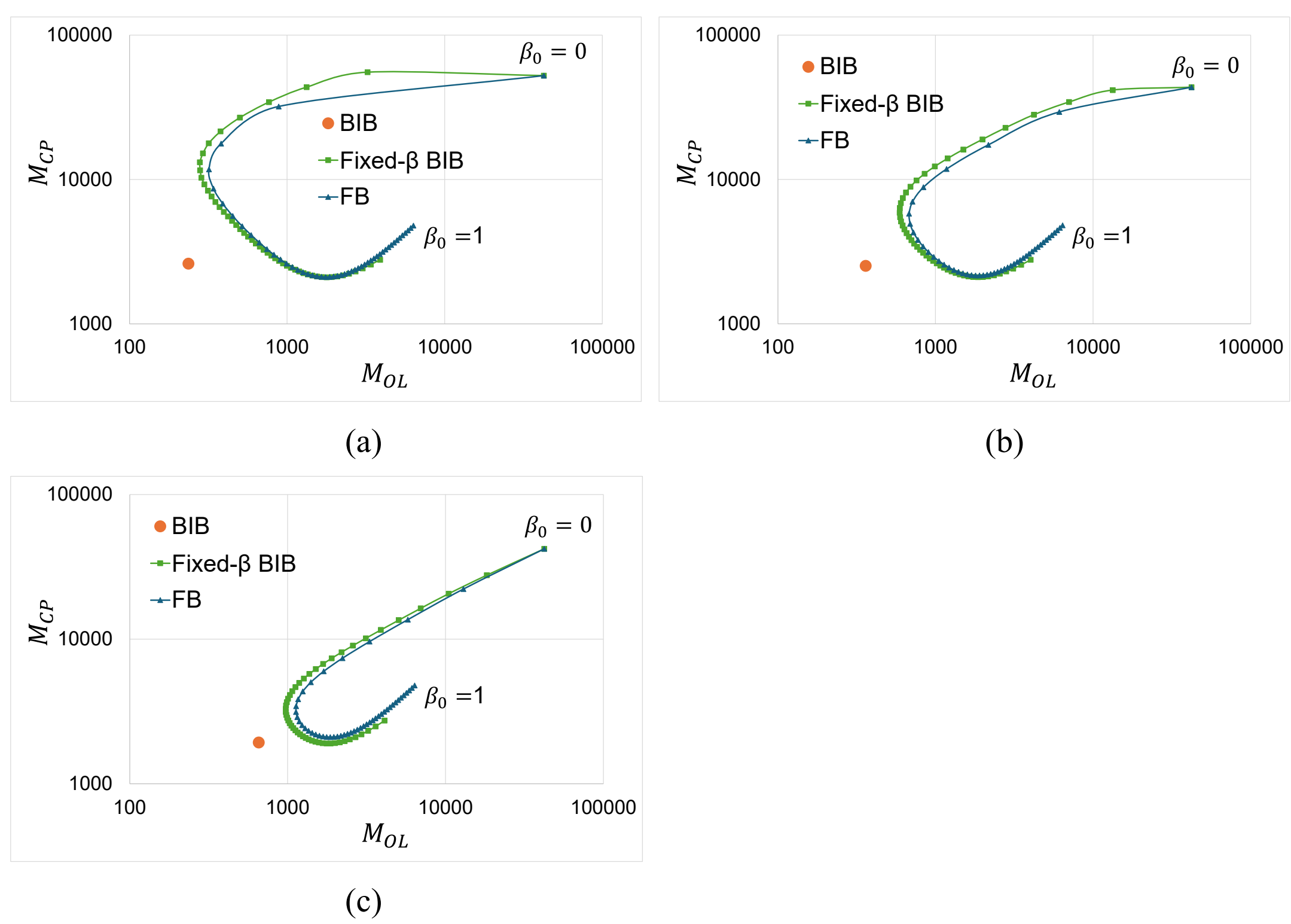

**Fig. 5** Relationship between tracking performance and stability for the BIB, fixed-$\beta$ BIB, and FB agents. For the fixed-$\beta$ BIB and FB agents, $\beta_0 \in [0,1]$ was scanned in increments of 0.02. For all three agents, the initial likelihood variance was set to the true value of the generative process, $R_0 = \sigma^2 = 100$. (a) $H^* = p_o^* = 0.001$. (b) $H^* = p_o^* = 0.01$. (c) $H^* = p_o^* = 0.05$. In each panel, the vertical axis represents the cumulative post-changepoint MSE, $M_{CP}$, and the horizontal axis represents the cumulative post-outlier MSE, $M_{OL}$. Both axes are on a logarithmic scale

## 4.4 Overall performance

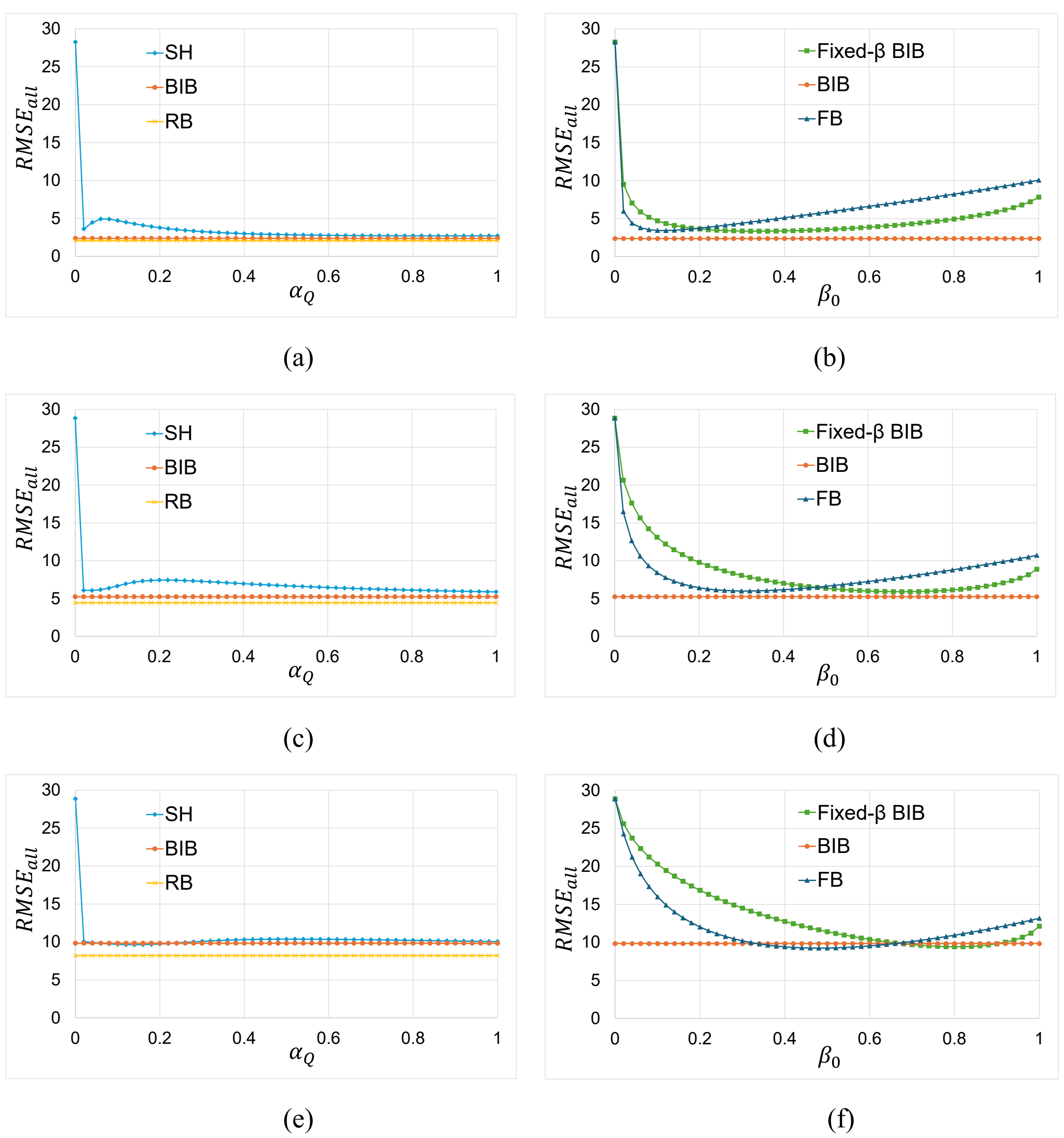

**Fig. 6** Overall RMSE for the BIB, RB, SH, fixed-$\beta$ BIB, and FB agents. For SH, $\alpha_Q \in [0,1]$ was scanned in increments of 0.02, whereas for fixed-$\beta$ BIB and FB, $\beta_0 \in [0,1]$ was scanned in increments of 0.02. BIB is shown as a horizontal line in both columns, whereas RB is shown as a horizontal line in the left column, because their performance does not depend on the parameter shown on the respective horizontal axis. For all five agents, the initial likelihood variance was set to the true value of the generative process, $R_0 = \sigma^2 = 100$. The left column (a, c, e) shows the SH scan over $\alpha_Q$, and the right column (b, d, f) shows the scans over $\beta_0$ for fixed-$\beta$ BIB and FB. (a, b) $H^* = p_o^* = 0.001$. (c, d) $H^* = p_o^* = 0.01$. (e, f) $H^* = p_o^* = 0.05$

Fig. 6 shows the overall RMSE under the three conditions $H^* = p_o^* \in \{0.001, 0.01, 0.05\}$. For each condition, the SH scan over $\alpha_Q$ is shown in the left column, whereas the scans over $\beta_0$ for fixed-$\beta$ BIB and FB are shown in the right column. BIB is overlaid as a horizontal line in both columns, whereas RB is shown as a horizontal line in the left column.

The overall RMSE of the RB was lower than that of the BIB under all conditions. The RB is provided with the true changepoint rate, outlier rate, and observation-noise variance. As shown in Figs. 1 and 2, RB adopted a higher learning rate than BIB at the event time, resulting in a smaller estimation error immediately after a changepoint. In contrast, because of its delayed-application mechanism, BIB maintained a relatively low learning rate immediately after the changepoint, leading to a larger estimation error than RB at $\tau = 0$.

When BIB was compared with SH, fixed-$\beta$ BIB, and FB, the horizontal line for BIB lay below the minima of the SH scan curve and the fixed-$\beta$ BIB and FB scan curves under the $H^* = p_o^* = 0.001$ and $H^* = p_o^* = 0.01$ conditions. By contrast, under the $H^* = p_o^* = 0.05$ condition, some parameter regions of the SH, fixed-$\beta$ BIB, and FB scan curves fell below the BIB horizontal line.

# 5 Discussion

In this study, we formulated the updating process of the continuous BIB models proposed by Shinohara et al. (2026) as two distinct variational processes: belief updating (B-step) and likelihood updating (L-step). The nullification strength $\beta$ shared by the two processes was determined endogenously to minimize the predictive surprise of the current observation under the candidate post-update predictive distribution. In the one-dimensional Gaussian setting with $\beta > 0$, the belief variance and the likelihood variance of the candidate update are expanded by a common factor relative to standard Bayesian updating, so that their post-update ratio is independent of $\beta$. This property allows BIB to temporarily weaken the constraints imposed by the existing predictive structure without immediately attributing a large prediction error to either a change in the latent state or an anomaly in the observation process.

In the simulations, BIB showed a favorable trade-off between tracking after changepoints and stability after outliers compared with the Sage–Husa (SH) model. Moreover, although BIB does not explicitly represent changepoints and outliers as distinct causal hypotheses, the learning rate profiles following the two types of events diverged in response to subsequent observations. In contrast, RB, which was provided with the true changepoint probability, outlier probability, and observation-noise variance, outperformed BIB in tracking after changepoints but showed lower stability than BIB after outliers.

## 5.1 Relationship with the original BIB and related methods

BIB was proposed as a framework that combines standard Bayesian inference with inverse Bayesian inference, thereby updating not only the belief distribution, but also the likelihood structure (Gunji et al. 2017). In the original discrete BIB, the hypothesis with the lowest posterior probability is selected, and its likelihood is replaced with an empirical distribution obtained from recent observations. This framework has been applied to the modeling of adaptive behaviors, such as foraging, in which it has been shown to generate search patterns exhibiting temporal correlations and power-law statistics (Shinohara et al. 2022). In repeated games, inverse Bayesian updating has been reported to generate heavy-tailed statistics for internal states (Sasai and Gunji 2026).

In the present study, the L-step was not a direct continuous extension of the original BIB. While the original BIB discontinuously replaces the likelihood of a particular hypothesis with an empirical distribution, this study continuously relativizes the conditional likelihoods associated with all the latent states by referring to the pre-update predictive distribution. Nevertheless, the two approaches share a common functional principle; they relativize the likelihood structure with respect to a reference distribution that is not conditioned on a particular latent state or hypothesis, thereby reducing excessive concentration on a specific hypothesis.

Insofar as both the belief distribution and model-side quantities are updated iteratively, BIB has structural similarities to the expectation–maximization (EM) algorithm and its variational formulation (Dempster et al. 1977; Neal and Hinton 1998). Because it adjusts the effective weight that observations exert on belief updating, BIB is also related to precision weighting under the free-energy principle (Feldman and Friston 2010), generalized Bayesian inference (Bissiri et al. 2016), power likelihoods, fractional posteriors (Holmes and Walker 2017; Bhattacharya et al. 2019), SafeBayes (Grünwald and van Ommen 2017), and coarsening (Miller and Dunson 2019). However, BIB differs from these approaches not simply in adjusting the effective influence of observations, but in how the likelihood structure itself is modified. Variational Bayesian inference and EM update the posterior distributions or model parameters to better explain the data under a generative model, whereas precision weighting and generalized Bayesian approaches adjust the effective weights of the prediction errors or likelihoods. By contrast, the L-step in BIB relativizes the likelihood in the opposite direction with reference to the pre-update predictive distribution, thereby weakening the constraints imposed by the predictive structure formed through the history of observations.

This distinction becomes clear when BIB is compared to a simple power-based weakening of likelihood. Suppose that the B-step is kept the same and the likelihood is updated according to $\tilde{l}_{t+1}^{\mathrm{pow}}(o \mid s; \beta) \propto l_t(o \mid s)^{1-\beta}$.

In the one-dimensional Gaussian setting, $\tilde{R}_{t+1}^{\mathrm{pow}}(\beta) = \frac{R_t}{1-\beta}$, whereas the B-step gives $\tilde{P}_{t+1}(\beta) = \frac{P_t R_t}{P_t + (1-\beta) R_t}$.

Their variance ratio is therefore $\frac{\tilde{R}^{\text{pow}}_{t+1}(\beta)}{\tilde{P}_{t+1}(\beta)} = \frac{1}{1-\beta} + \frac{R_t}{P_t}$, which depends explicitly on $\beta$. In contrast, under the BIB update, $\frac{\tilde{R}_{t+1}(\beta)}{\tilde{P}_{t+1}(\beta)} = 1 + \frac{R_t}{P_t}$, which is independent of $\beta$.

Thus, the BIB simultaneously reduces the absolute precision of both the belief and observation sides without changing the relative uncertainty.

As a framework that explicitly decomposes prediction error into state-side variability and observation-side stochasticity, Piray and Daw (2021) proposed a model that jointly estimates latent-state volatility and observation stochasticity as separate latent variables and adjusts the learning rate accordingly. Piray and Daw (2024) demonstrate that humans can infer these two forms of uncertainty separately and modulate their learning rates accordingly.

In contrast, BIB does not allocate the contribution of the prediction error between these two sources. Instead, it determines a single scalar quantity, $\beta$, that specifies the extent to which the constraints imposed by the predictive structure as a whole should be weakened. Therefore, BIB does not explicitly separate the causes of prediction error but instead leaves the differentiation of subsequent adaptive responses to later observations.

## 5.2 Nullification as suspension of causal attribution

Standard Bayesian updating can be viewed as comprising two components: a constraint that favors retaining the prior belief, and a likelihood-driven force that links observations to latent states. In the present study, belief forgetting weakened the former, whereas the expansion of the likelihood variance reduced the certainty of the latter.

From the perspective of causal attribution, attributing a prediction error to a changepoint corresponds to weakening the constraint imposed by the prior belief, whereas attributing it to an outlier corresponds to weakening the observation–state relationship. Functionally, the former facilitates observation tracking, whereas the latter suppresses observation-driven updating.

By contrast, nullification in BIB weakens both constraints simultaneously through a common $\beta$. In the one-dimensional Gaussian setting, the belief variance and likelihood variance are enlarged by the same factor relative to the standard Bayesian update, leaving their ratio unchanged. Thus, at least at the variance ratio level, nullification can be interpreted as an operation that suspends the preferential attribution of a prediction error to either the state or observation side.

The role of the L-step is not merely to reduce the influence of observations according to the selected $\beta$. Because the B-step and L-step are constructed under a common $\beta$ that is determined by minimizing predictive surprise under the candidate predictive distribution obtained by applying both steps, $\beta$ represents not simply the degree of belief forgetting but the extent to which the constraints imposed by the predictive structure as a whole are weakened. In this sense, the L-step is essential for defining the meaning of the nullification strength in a way that does not privilege either form of causal attribution.

## 5.3 Context-dependent learning through subsequent observations

Nassar et al. (2019) showed that, in a predictive task in which changepoint and outlier conditions were experienced in separate blocks, the learning rate increased with surprise in the changepoint condition but decreased in the outlier condition. This finding indicates that the amount of learning depends not only on the magnitude of the prediction error but also on the statistical context in which that error occurs.

In contrast, in the present study, changepoints and outliers were intermingled within the same sequence, and the observation distributions associated with the two event types were set to be identical. Consequently, the cause of an event cannot be determined from a single observation occurring in the event. Therefore, it is important not to select a cause immediately at the time of the event but rather to produce a common initial response and subsequently differentiate that response according to later observations.

In BIB, the nullification strength calculated from a large prediction error is not applied at the same time point but is retained as a candidate for the next time step. Consequently, at the event time ($\tau = 0$), the learning rates following changepoints and outliers are nearly identical. Thereafter, if observations consistent with the new state continue to occur after a changepoint, tracking is facilitated by a positive $\beta$. By contrast, if observations return to the vicinity of the original latent mean after an outlier, the standard Bayesian candidate, in which $\beta = 0$ and the likelihood variance is returned to its baseline value ($R_0$), is more likely to be selected. Thus, the asymmetry between the responses following the two event types does not arise from the causal classification of the event itself but from the interaction between delayed nullification and subsequent observations. As shown by the internal variable profiles in Fig. 3, this differentiation arose primarily from the subsequent adoption or rejection of the nullification candidate rather than from a systematic difference in the nullification strength conditional on retention.

This reset mechanism differs from belief restarting in changepoint models. Bayesian online changepoint detection and related models explicitly introduce quantities such as a changepoint hypothesis, run length, and hazard rate (Adams and MacKay 2007; Wilson et al. 2010; Nassar et al. 2010). In contrast, BIB does not maintain changepoints or outliers as latent hypotheses. Instead, BIB uses the predictive consistency with subsequent observations to determine whether nullification should be maintained or released. Responses that were initially common at the time of the event become differentiated over time.

The RB model explicitly maintains causal hypotheses corresponding to changepoints, nominal observations, and outliers. However, in the present study, the changepoint and outlier probabilities were set to be equal, and the observation distributions for the two event types were identical. Consequently, no clear difference emerged in the learning rate immediately after the events, and the subsequent differentiation of the mean learning rate profiles was limited. Under symmetric conditions, explicitly representing alternative causal hypotheses does not guarantee differentiated responses.

The SH model used in the present study also produced differentiated responses following the two event types through adaptation of process-noise variance ($Q_t$). However, this differentiation emerged more slowly than in BIB and depended on the adaptation rate ($\alpha_Q$). Although the delayed application employed by BIB does not enable immediate discrimination, it generates a differentiation of responses based on subsequent observations at a relatively early stage.

The learning-rate profiles characterize the internal adaptive responses of the models, whereas the MSE profiles indicate how those responses affect the estimation accuracy. In BIB, the nullification candidate generated by the event itself has not yet been applied at the event time, so the MSE immediately after a changepoint tends to be relatively large. From the next time step onward, however, tracking is accelerated. By contrast, following an outlier, the persistence of its influence is suppressed through the reset mechanism.

## 5.4 Trade-off between tracking and stability and overall performance

In the trade-off between tracking and stability, the BIB was nondominated by the SH parameter sweep across all event-rate conditions examined. In contrast, RB outperformed BIB in post-changepoint tracking under all conditions but performed worse than BIB in post-outlier stability. However, RB was provided with the true changepoint rate, outlier rate, and observation-noise variance; therefore, its advantage partly reflects the benefits of this oracle information.

In the comparison among BIB, fixed-$\beta$ BIB, and the FB agent, the two sweeps lay close to each other near their most favorable settings, and their minima of overall RMSE differed only marginally, with FB slightly lower under $H^* = p_o^* = 0.05$. What distinguished the two agents was their behavior at large $\beta_0$: as $\beta_0$ approached 1, the overall RMSE of FB increased markedly, whereas that of fixed-$\beta$ BIB increased far less (Fig. 6(b), (d), (f)), and correspondingly the FB sweep extended to larger $M_{OL}$ in the cumulative MSE plane (Fig. 5). This pattern is consistent with the mechanism described below: without the L-step, the steady-state learning rate is fixed at $\beta_0$, so that a large forgetting strength translates directly into persistent over-updating, whereas the combination of the L-step and reset bounds this effect. Across all three event-rate conditions for either agent, no scanned value of $\beta_0$ achieved smaller $M_{CP}$ and $M_{OL}$ values than BIB; under $H^* = p_o^* = 0.05$, BIB dominated the entire FB sweep. Under the comparison conditions used in this study, these results suggest roles for both the control mechanism consisting of likelihood-variance expansion and reset and the endogenous determination of the nullification strength.

The L-step and reset operate as an integrated control mechanism. Without the L-step, the likelihood variance remains at $R_0$, so that the reset reduces to switching nullification off for a single time step and no longer restores the certainty of the observation–state correspondence. Conversely, if an L-step is introduced without resetting, $\frac{R_{t+1}}{P_{t+1}} = 1 + \frac{R_t}{P_t}$ holds independently of $\beta$. Consequently, $R_t/P_t$ increases with repeated updates, which reduces the learning rate.

With the B-step alone, the increase in belief variance caused by belief forgetting remains, so the increase in the learning rate persists. In contrast, the L-step simultaneously expands the likelihood variance, thereby preventing the learning-rate increase from persisting unconditionally, whereas the reset returns the likelihood variance to $R_0$, temporally localizing the effect of nullification. Thus, for an elevated learning rate to be maintained, the nullification-continuation candidate must continue to be selected based on subsequent observations. The performance of BIB can therefore be attributed not only to belief forgetting but also to the combination of likelihood-variance control and the endogenous determination of nullification strength.

However, the model ranking based on the trade-off evaluated using the cumulative post-event MSE did not always coincide with that based on the overall RMSE computed over all time points. The cumulative MSE evaluates the conditional recovery process following an event, whereas the overall RMSE evaluates all time points in a time-weighted manner, including periods of nominal observations. In BIB, nullification is not applied until the next time step, making its immediate response at the event time less favorable than that of models that react immediately. However, during periods of nominal observation, BIB avoids excessive nullification, and after an event, recovers either tracking or stability according to subsequent observations. Consequently, as the event rate increases, the initial cost associated with the delayed application is repeatedly incurred, reducing BIB's advantage in terms of the overall RMSE.

Therefore, the difference between the cumulative MSE and overall RMSE can be understood as reflecting differences in the timescale of evaluation and the weight of the initial cost of delayed application. The advantage of BIB lies not in immediately classifying the cause of an event when it occurs but in preserving the estimation efficiency during nominal periods, while subsequently differentiating between tracking and stability based on later observations.

## 5.5 Limitations and future directions

This study has several limitations.

First, environmental conditions were limited. In this study, the changepoint and outlier rates were set equal, and their relative prevalence, observation-noise variance, event magnitude, and distributional form were not systematically manipulated. Future work should examine the behavior of nullification and resetting under a broader range of conditions, including environments in which either changepoints or outliers predominate. In addition, our comparison with hierarchical models that separately estimate volatility and stochasticity (Piray and Daw 2021, 2024) remains at the conceptual level. A fair numerical comparison requires a careful consideration of the correspondence between the respective generative processes.

Second, this study assumed a one-dimensional Gaussian distribution. In the multivariate Gaussian case, it is necessary to derive the conditions under which the matrix relationships corresponding to the equal-factor expansion and variance-ratio preservation found in the one-dimensional case hold. The properties of the L-step under non-Gaussian distributions and nonlinear observation models are yet to be investigated.

Third, although the B-step and L-step were functionally separated in the present study, we did not examine a fully coupled formulation in which the L-step acted directly on belief updating. It is necessary to determine the conditions under which the symmetric variance expansion of the BIB is preserved in such a coupled model.

Fourth, because the post-event evaluation window was defined as $\tau = 0, \ldots, 100$ and truncated at the occurrence of the next event, the number of valid events decreases as $\tau$ increases. Moreover, estimates at later time points were conditioned on relatively long event-free sequences before the next event. Therefore, the effects of the evaluation window and truncation rule on temporal profiles and cumulative MSE should be examined in future studies.

Fifth, the delayed application of $\beta$, the reset rule, and the determination of $\beta$ based on local predictive surprise are specific implementation choices adopted in the present study. A delayed application reflects the design principle of not immediately committing to a causal interpretation based on a single observation, but this comes at the cost of reduced responsiveness at the time of the event. Indeed, under high event-rate conditions, some settings of SH, fixed-$\beta$ BIB, or FB outperformed BIB in terms of overall RMSE. Future work should therefore compare BIB with variants using immediate application, alternative reset rules, or a nonzero baseline $\beta$. In addition, although the BIB operates without externally supplied event rates, it depends on the initial likelihood variance $R_0$. Thus, while preserving the design principle of avoiding prior specification of environmental event statistics, future work should examine adaptive estimation of the baseline likelihood variance as well as alternative mechanisms for determining the nullification strength that balance local predictive surprise against long-term performance.

Sixth, the present results are based on simulations. It remains unknown whether a common initial response that defers causal attribution, followed by subsequent differentiation of learning rate, also occurs in human or biological inferences. This should be tested experimentally using behavioral paradigms in which changepoints and outliers are interleaved within the same sequence and the observation distributions at event times are controlled.

Taken together, the central feature of BIB is that it does not immediately determine the cause of large prediction errors. Instead, it simultaneously relaxes constraints on both the belief and likelihood sides and subsequently differentiates its adaptive response according to later observations. Future studies should determine whether this design principle is effective across more general environments, distributions, model settings, and biological inferences.

# 6 Conclusion

BIB does not immediately classify a large prediction error as either a changepoint or an outlier. Instead, it temporarily relaxes both belief- and likelihood-side constraints and allows subsequent observations to determine whether this relaxed state should be maintained or released.

This delayed differentiation provides a distinct adaptive principle: uncertainty about the cause of prediction error is handled not by immediately selecting among explicit causal hypotheses, but by temporarily weakening the existing predictive structure and allowing later observations to shape the subsequent response.

In the present simulations, this mechanism generated an initially common response to changepoints and outliers followed by rapid event-dependent differentiation, while providing a favorable trade-off between post-changepoint tracking and post-outlier stability. These results suggest that temporarily suspending causal attribution may provide a useful principle for adaptive inference under ambiguous prediction errors.

# Appendix

## Variational derivation of the standard Bayesian update

We derive the variational form of the standard Bayesian update defined in Section 2.1.1 of the main text.

Let $q_t(s)$ denote the belief before updating and $l_t(o_t \mid s)$ the likelihood. For a candidate belief $q(s)$, define

$$[q] = \mathbb{E}_{q(s)}\left[-\log l_t\left(o_t|s\right)\right] + D_{KL}\left(q(s) \,||\, q_t(s)\right). \tag{A1}$$

The predictive distribution before updating is $p_t(o_t) = \int q_t(s) l_t(o_t|s) ds$ and the Bayesian posterior is given by $p_t(s|o_t) = \frac{q_t(s) l_t(o_t|s)}{p_t(o_t)}$.

The KL divergence between the candidate belief $q(s)$ and the Bayesian posterior is

$$\begin{aligned} D_{KL}\left(q(s) \,||\, p_t(s|o_t)\right) &= \int q(s) \log \frac{q(s)}{p_t(s|o_t)} ds \\ &= \int q(s) \log \frac{q(s)}{q_t(s)} ds - \int q(s) \log l_t(o_t|s) ds + \log p_t(o_t). \end{aligned} \tag{A2}$$

Therefore,

$$D_{KL}\left(q(s) \,||\, p_t(s|o_t)\right) = F[q] + \log p_t(o_t), \tag{A3}$$

and hence

$$F[q] = -\log p_t(o_t) + D_{KL}\left(q(s) \,||\, p_t(s|o_t)\right) \tag{A4}$$

Because the KL divergence is non-negative, $F[q] \geq -\log p_t(o_t)$, with equality when $q(s) = p_t(s \mid o_t)$.

Therefore,

$$q_{t+1}(s) = \arg\min_{q(s)} F[q] = p_t(s|o_t), \tag{A5}$$

showing that minimizing this variational objective is equivalent to the standard Bayesian update.

## Belief update (B-step)

We consider the B-step objective function defined in Section 2.1.2 of the main text. For candidate belief $q(s)$ given an observation $o_t$, let

$$\begin{aligned} \tilde{q}_{t+1}(s;\beta) &= \arg\min_{q(s)} F_B[q;\beta], \\ F_B[q;\beta] &= \mathbb{E}_{q(s)}\left[-\log l_t(o_t|s)\right] + (1-\beta) D_{KL}\left(q(s) \,\|\, q_t(s)\right) + \beta D_{KL}\left(q(s) \,\|\, u(s)\right). \end{aligned} \tag{A6}$$

Here, $0 \leq \beta \leq 1$, and $u(s)$ is a reference probability density for forgetting.

Introducing a Lagrange multiplier $\lambda$ under the normalization constraint $\int q(s)ds = 1$, we obtain

$$L[q] = \int q(s)\left[-\log l_t(o_t|s)\right]ds + (1-\beta)\int q(s)\log\frac{q(s)}{q_t(s)}ds + \beta\int q(s)\log\frac{q(s)}{u(s)}ds + \lambda\left(\int q(s)ds - 1\right). \tag{A7}$$

Taking the functional derivative with respect to $q(s)$ gives

$$\frac{\delta L}{\delta q(s)} = -\log l_t(o_t|s) + (1-\beta)\left[\log\frac{q(s)}{q_t(s)} + 1\right] + \beta\left[\log\frac{q(s)}{u(s)} + 1\right] + \lambda. \tag{A8}$$

Using the stationarity condition $\frac{\delta L}{\delta q(s)} = 0$ and $(1-\beta) + \beta = 1$, we obtain

$$\log q(s) = (1-\beta)\log q_t(s) + \beta\log u(s) + \log l_t(o_t|s) + \text{const.} \tag{A9}$$

Therefore,

$$\tilde{q}_{t+1}(s;\beta) \propto q_t(s)^{1-\beta} u(s)^{\beta} l_t(o_t|s). \tag{A10}$$

For a finite or bounded state space, if $u(s)$ is considered uniform, its $s$-independent factor can be absorbed into the normalization constant, yielding

$$\tilde{q}_{t+1}(s;\beta) \propto q_t(s)^{1-\beta} l_t(o_t|s). \tag{A11}$$

Because a uniform probability density cannot be normalized on an unbounded state space, the "flat reference" in the Gaussian setting considered in this study may instead be interpreted by introducing, for example, a sufficiently broad Gaussian distribution, $u_\Lambda(s) = \mathcal{N}(s; \mu_u, \Lambda)$, and taking the limit $\Lambda \to \infty$. In this limit, the $s$-dependent contribution of $u_\Lambda(s)^\beta$ vanishes, recovering the expression for the flat-reference case above.

When $\beta = 0$, $\tilde{q}_{t+1}(s; 0) \propto q_t(s) l_t(o_t \mid s)$, which coincides with the standard Bayesian update. In contrast, when $\beta > 0$, the contribution of the prior belief $q_t(s)$ is weakened, thereby reducing the constraint imposed by the prior belief.

## B-step in the Gaussian setting

Let the belief distribution and likelihood be

$$\begin{aligned} q_t(s) &= \mathcal{N}(s; m_t, P_t), \\ l_t(o_t|s) &= \mathcal{N}(o_t; s, R_t). \end{aligned} \tag{A12}$$

Under a flat reference, the B-step is given by $\tilde{q}_{t+1}(s;\beta) \propto q_t(s)^{1-\beta} l_t(o_t \mid s)$.

Focusing on the exponential term of the Gaussian distribution, $q_t(s)^{1-\beta} \propto \exp\left[-\frac{1-\beta}{2P_t}(s-m_t)^2\right]$,

and $l_t(o_t \mid s) \propto \exp\left[-\frac{1}{2R_t}(o_t-s)^2\right]$. Therefore,

$$\tilde{q}_{t+1}(s;\beta) \propto \exp\left[-\frac{1}{2}\left\{\frac{1-\beta}{P_t}(s-m_t)^2 + \frac{1}{R_t}(o_t-s)^2\right\}\right]. \qquad \text{(A13)}$$

Collecting the quadratic terms in $s$, the precision of the candidate posterior distribution is

$\frac{1}{\tilde{P}_{t+1}(\beta)} = \frac{1-\beta}{P_t} + \frac{1}{R_t}$. Hence,

$$\tilde{P}_{t+1}(\beta) = \frac{P_t R_t}{P_t + (1-\beta)R_t} \qquad \text{(A14)}$$

is obtained.

From the linear term, the candidate posterior mean is

$$\tilde{m}_{t+1}(\beta) = \tilde{P}_{t+1}(\beta)\left[\frac{1-\beta}{P_t}m_t + \frac{1}{R_t}o_t\right]. \qquad \text{(A15)}$$

Substituting the expression for $\tilde{P}_{t+1}(\beta)$ gives

$$\tilde{m}_{t+1}(\beta) = \frac{(1-\beta)R_t m_t + P_t o_t}{P_t + (1-\beta)R_t}. \qquad \text{(A16)}$$

Defining the prediction error as $\delta_t = o_t - m_t$, this can be written as

$$\tilde{m}_{t+1}(\beta) = m_t + \frac{P_t}{P_t + (1-\beta)R_t}\delta_t. \qquad \text{(A17)}$$

Thus, defining the effective Kalman gain in the B-step as

$$\tilde{K}_t(\beta) = \frac{P_t}{P_t + (1-\beta)R_t}, \qquad \text{(A18)}$$

we have

$$\tilde{m}_{t+1}(\beta) = m_t + \tilde{K}_t(\beta)\delta_t. \qquad \text{(A19)}$$

When $\beta = 0$, $\widetilde{K}_t(0) = \frac{P_t}{P_t+R_t}$, and $\tilde{P}_{t+1}(0) = \frac{P_t R_t}{P_t+R_t}$, which coincide with the standard Gaussian Bayesian update.

In contrast, when $\beta > 0$, $P_t + (1-\beta)R_t < P_t + R_t$, and therefore $\widetilde{K}_t(\beta) > \widetilde{K}_t(0)$ and $\tilde{P}_{t+1}(\beta) > \tilde{P}_{t+1}(0)$.

Thus, by weakening the constraint imposed by the prior belief, the B-step increases the learning rate for the observation, while simultaneously increasing the post-update belief variance relative to that under the standard Bayesian update.

## Likelihood update (L-step)

Next, we consider the L-step objective function defined in Section 2.1.3. Pre-update predictive distribution $p_t(o) = \int q_t(s) l_t(o \mid s) ds$ is held fixed during minimization of the L-step objective. The objective function is

$$F_L[l;\beta] = \mathbb{E}_{q_t(s)}\left[D_{KL}\left(l(\cdot|s) \,||\, l_t(\cdot|s)\right)\right] + \beta \mathbb{E}_{q_t(s)} \mathbb{E}_{l(\cdot|s)}\left[\log p_t(o)\right]. \quad \text{(A20)}$$

For each state satisfying $q_t(s) > 0$, the positive weight $q_t(s)$ does not affect the minimization solution. Therefore, it is sufficient to consider that

$$\tilde{l}_{t+1}(o|s;\beta) = \arg\min_{l(\cdot|s)} \left\{D_{KL}\left(l(\cdot|s) \,||\, l_t(\cdot|s)\right) + \beta \mathbb{E}_{l(\cdot|s)}\left[\log p_t(o)\right]\right\}. \quad \text{(A21)}$$

Introducing a Lagrange multiplier $\lambda_s$ for the normalization constraint $\int l(o \mid s) do = 1$, gives

$$L_s[l] = \int l(o|s) \log \frac{l(o|s)}{l_t(o|s)} do + \beta \int l(o|s) \log p_t(o) do + \lambda_s \left(\int l(o|s) do - 1\right). \quad \text{(A22)}$$

Taking the functional derivative with respect to $l(o \mid s)$ yields $\frac{\delta L_s}{\delta l(o|s)} = \log \frac{l(o|s)}{l_t(o|s)} + 1 +$

$\beta \log p_t(o) + \lambda_s$. The stationarity condition therefore gives $\log l(o \mid s) = \log l_t(o \mid s) -$

$\beta \log p_t(o) + \text{const}(s)$. Hence,

$$l(o|s) \propto l_t(o|s)\, p_t(o)^{-\beta}. \tag{A23}$$

Defining the normalization constant as $Z_t(s;\beta) = \int l_t(o \mid s) p_t(o)^{-\beta} do$, we obtain

$$\tilde{l}_{t+1}(o|s;\beta) = \frac{l_t(o|s)\, p_t(o)^{-\beta}}{Z_t(s;\beta)}. \tag{A24}$$

However, for this solution to exist as a probability density, $Z_t(s;\beta) < \infty$.

This update weakens the regions in which the predictive density $p_t(o)$ is high, relative to the regions in which it is low. In general, this operation may alter the location, variance, and overall shape of likelihood. In this study, to preserve the direct correspondence $o \simeq s$ between the observation and latent states, we restrict the candidate likelihood to a fixed-center Gaussian family.

## L-step within the fixed-center Gaussian family

We restrict the candidate likelihood to the center-fixed Gaussian family $l_V(o \mid s) =$

$\mathcal{N}(o; s, V), \quad V > 0$, and optimize only the common variance $V$.

In this case,

$$D_{KL}\left(\mathcal{N}(s,V) \,\|\, \mathcal{N}(s,R_t)\right) = \frac{1}{2}\left[\frac{V}{R_t} - 1 - \log \frac{V}{R_t}\right]. \tag{A25}$$

Furthermore, $\log p_t(o) = -\frac{1}{2}\log[2\pi(P_t + R_t)] - \frac{(o-m_t)^2}{2(P_t+R_t)}$.

Because $o \sim \mathcal{N}(s, V)$ and $s \sim q_t(s)$, $\mathbb{E}_{q_t(s)}\mathbb{E}_{\mathcal{N}(o;s,V)}[(o - m_t)^2] = V + P_t$. Therefore,

$$F_L(V;\beta) = \frac{1}{2}\left[\frac{V}{R_t} - 1 - \log\frac{V}{R_t}\right] - \frac{\beta}{2}\log\left[2\pi(P_t + R_t)\right] - \frac{\beta(V + P_t)}{2(P_t + R_t)}. \quad \text{(A26)}$$

Collecting terms independent of $V$ into a constant gives

$$F_L(V;\beta) = \frac{1}{2}\left[\frac{V}{R_t} - 1 - \log\frac{V}{R_t}\right] - \frac{\beta V}{2(P_t + R_t)} + \text{const.} \quad \text{(A27)}$$

Differentiating with respect to $V$ gives $\frac{\partial F_L}{\partial V} = \frac{1}{2}\left[\frac{1}{R_t} - \frac{1}{V} - \frac{\beta}{P_t + R_t}\right]$.

Thus, the stationarity condition $\frac{\partial F_L}{\partial V} = 0$ yields $\frac{1}{V} = \frac{1}{R_t} - \frac{\beta}{P_t + R_t}$.

Hence, $V = \left(\frac{1}{R_t} - \frac{\beta}{P_t + R_t}\right)^{-1}$, and therefore

$$\tilde{R}_{t+1}(\beta) = \frac{R_t(P_t + R_t)}{P_t + (1-\beta)R_t} \quad \text{(A28)}$$

Moreover, $\frac{\partial^2 F_L}{\partial V^2} = \frac{1}{2V^2} > 0$.

Therefore, this stationary solution is a unique minimum within the fixed-center Gaussian family.

For $P_t > 0$, $R_t > 0$, and $0 \leq \beta \leq 1$, $P_t + (1 - \beta)R_t > 0$ always holds.

When $\beta = 0$, $\tilde{R}_{t+1}(0) = R_t$ and hence the likelihood remains unchanged. In contrast, when $\beta > 0$, $\tilde{R}_{t+1}(\beta) > R_t$. Thus, the L-step expands the likelihood variance, thereby reducing the certainty of the correspondence between the observation and latent state.

## Endogenous determination of the nullification strength $\boldsymbol{\beta}$ in BIB

The nullification strength was determined to minimize the predictive surprise of the current observation $o_t$ under the candidate post-update predictive distribution obtained by applying both the B- and L-steps.

The candidate belief obtained by the B-step is $\tilde{q}_{t+1}(s;\beta) = \mathcal{N}\left(s; \tilde{m}_{t+1}(\beta), \tilde{P}_{t+1}(\beta)\right)$, where $\tilde{P}_{t+1}(\beta) = \frac{P_t R_t}{P_t + (1-\beta)R_t}$, and $\tilde{m}_{t+1}(\beta) = m_t + \frac{P_t}{P_t + (1-\beta)R_t}\delta_t$.

Meanwhile, the L-step within the center-fixed Gaussian family gives $\tilde{R}_{t+1}(\beta) = \frac{R_t(P_t + R_t)}{P_t + (1-\beta)R_t}$.

The candidate post-update predictive distribution is therefore $\tilde{p}_{t+1}(o;\beta) = \mathcal{N}\left(o; \tilde{m}_{t+1}(\beta), \tilde{P}_{t+1}(\beta) + \tilde{R}_{t+1}(\beta)\right)$.

Let $x = 1 - \beta,\ d = P_t + xR_t$. Then,

$$\begin{aligned} \tilde{m}_{t+1} &= m_t + \frac{P_t}{d}\delta_t. \\ \tilde{P}_{t+1} &= \frac{P_t R_t}{d}, \\ \tilde{R}_{t+1} &= \frac{R_t\left(P_t + R_t\right)}{d}. \end{aligned} \tag{A29}$$

Thus, the candidate predictive variance is $\tilde{P}_{t+1} + \tilde{R}_{t+1} = \frac{R_t(2P_t + R_t)}{d}$.

Furthermore, $o_t - \tilde{m}_{t+1} = \delta_t - \frac{P_t}{d}\delta_t = \frac{xR_t}{d}\delta_t$.

Therefore, the log density of the candidate predictive distribution evaluated at $o_t$ is

$$\log \tilde{p}_{t+1}\left(o_t; x\right) = \frac{1}{2}\log\left(P_t + xR_t\right) - \frac{x^2 R_t \delta_t^2}{2\left(2P_t + R_t\right)\left(P_t + xR_t\right)} + C, \tag{A30}$$

where $C$ is independent of $x$.

Differentiating with respect to $x$ yields

$$\frac{\partial}{\partial x}\log \tilde{p}_{t+1}(o_t;x) = \frac{R_t}{2(2P_t+R_t)(P_t+xR_t)^2}G_t(x),$$
$$G_t(x) = (P_t+xR_t)(2P_t+R_t) - \delta_t^2 x(2P_t+xR_t). \tag{A31}$$

The stationarity condition is $G_t(x) = 0$, which can be rearranged as

$$R_t\delta_t^2 x^2 + \left[2P_t\delta_t^2 - R_t(2P_t+R_t)\right]x - P_t(2P_t+R_t) = 0. \tag{A32}$$

For $\delta_t \neq 0$, the product of the two roots is $-\frac{P_t(2P_t+R_t)}{R_t\delta_t^2} < 0$.

Thus, there is a unique positive root. Denoting this positive root by $x^*$, we obtain:

$$x^* = \frac{R_t(2P_t+R_t) - 2P_t\delta_t^2 + \sqrt{R_t^2(2P_t+R_t)^2 + 4P_t^2\delta_t^4}}{2R_t\delta_t^2}. \tag{A33}$$

Furthermore, $G_t(0) = P_t(2P_t + R_t) > 0$, and the sign of $G_t(x)$ changes from positive to negative at the positive root, $x^*$. Therefore, $\log \dot{p}_{t+1}(o_t; x)$ increases up to $x^*$ and decreases thereafter. Thus, $x^*$ is a unique maximum in the positive domain.

Moreover, $G_t(1) = (2P_t + R_t)(P_t + R_t - \delta_t^2)$. Therefore, when $\delta_t^2 \leq P_t + R_t$, we have $x^* \geq 1$. In this case, the predictive density increases with $x$ over the admissible range $0 \leq x \leq 1$, and hence its maximum is attained at $x = 1$, that is, $\beta = 0$.

In contrast, when $\delta_t^2 > P_t + R_t$, we have $0 < x^* < 1$, and the maximum within the admissible range is attained at $x = x^*$. Therefore, $\beta^* = 1 - x^*$.

Accordingly, the candidate nullification strength retained for the next time step is

$$\tilde{\beta}_{t+1} = \begin{cases} 0, & \delta_t^2 \leq P_t + R_t, \\ 1 - \dfrac{R_t(2P_t+R_t) - 2P_t\delta_t^2 + \sqrt{R_t^2(2P_t+R_t)^2 + 4P_t^2\delta_t^4}}{2R_t\delta_t^2}, & \delta_t^2 > P_t + R_t. \end{cases} \tag{A34}$$

Thus, when $\frac{\delta_t^2}{P_t+R_t} \leq 1$, no positive nullification candidates were generated for the next time step. In contrast, when $\frac{\delta_t^2}{P_t+R_t} > 1$, we obtain $0 < \tilde{\beta}_{t+1} < 1$, and a positive nullification strength was retained as a candidate for the next time step.

The value $\tilde{\beta}_{t+1}$ obtained here is not immediately applied to the update associated with observation $o_t$. As described in Section 2.3 of the main text, it is retained as an update candidate for the next time step, and its adoption is determined through a reset decision based on the predictive surprise of the subsequent observation.

## Derivation of the posterior variance for the RB agent

We derived the posterior variance of the RB agent used in the main text.

At time $t$, let the posterior probabilities of the changepoint, nominal observation, and outlier hypotheses be denoted as $\Omega_t, \Gamma_t, O_t$, respectively. These satisfy $\Omega_t + \Gamma_t + O_t = 1$.

Under the changepoint hypothesis, the updated mean and variance are $m_{t+1}^{CP} = o_t, \quad P_{t+1}^{CP} = 0$.

Under the nominal-observation hypothesis, $m_{t+1}^{N} = m_t + \alpha_t\delta_t, \quad P_{t+1}^{N} = (1-\alpha_t)P_t$.

Under the outlier hypothesis, $m_{t+1}^{O} = m_t, \quad P_{t+1}^{O} = P_t$.

Here, $\delta_t = o_t - m_t$ is the prediction error, and $\alpha_t = \frac{P_t}{P_t+R_0}$ is the Kalman gain under the nominal-observation hypothesis.

The updated mean of the three-hypothesis mixture is $m_{t+1} = \Omega_t m_{t+1}^{CP} + \Gamma_t m_{t+1}^{N} + O_t m_{t+1}^{O}$.

Substituting the mean under each hypothesis gives

$$m_{t+1} = \Omega_t o_t + \Gamma_t \left( m_t + \alpha_t \delta_t \right) + O_t m_t. \tag{A35}$$

Using $o_t = m_t + \delta_t$ and $\Omega_t + \Gamma_t + O_t = 1$, we obtain

$$m_{t+1} = m_t + \left( \Omega_t + \Gamma_t \alpha_t \right) \delta_t. \tag{A36}$$

Thus, defining the effective learning rate as $K_t = \Omega_t + \Gamma_t \alpha_t$, the update can be written as $m_{t+1} = m_t + K_t \delta_t$.

Next, based on the law of total variance, the updated variance of the mixture distribution is

$$P_{t+1} = \sum_j w_j \left[ P_{t+1}^j + \left( m_{t+1}^j - m_{t+1} \right)^2 \right], \tag{A37}$$

where $j \in \{CP, N, O\}$, and $w_j \in \{\Omega_t, \Gamma_t, O_t\}$.

First, we compute the difference between the mean of each hypothesis and the mixture mean. Under the changepoint hypothesis,

$$m_{t+1}^{CP} - m_{t+1} = o_t - \left( m_t + K_t \delta_t \right) = \left( 1 - K_t \right) \delta_t. \tag{A38}$$

Under the nominal-observation hypothesis,

$$m_{t+1}^{N} - m_{t+1} = \left( m_t + \alpha_t \delta_t \right) - \left( m_t + K_t \delta_t \right) = \left( \alpha_t - K_t \right) \delta_t. \tag{A39}$$

Under the outlier hypothesis,

$$m_{t+1}^{O} - m_{t+1} = m_t - \left( m_t + K_t \delta_t \right) = -K_t \delta_t. \tag{A40}$$

Substituting the within-hypothesis variances and differences in means into the law of total variance yields $P_{t+1} = \Omega_t[0 + (1 - K_t)^2 \delta_t^2] + \Gamma_t[(1 - \alpha_t) P_t + (\alpha_t - K_t)^2 \delta_t^2] + O_t[P_t + K_t^2 \delta_t^2]$.

Rearranging gives

$$P_{t+1} = \Gamma_t \left(1-\alpha_t\right) P_t + O_t P_t + \left[\Omega_t \left(1-K_t\right)^2 + \Gamma_t \left(\alpha_t - K_t\right)^2 + O_t K_t^2\right] \delta_t^2 . \tag{A41}$$

The first two terms on the right-hand side represent the within-hypothesis variances weighted by their posterior probabilities, whereas the final term represents the between-hypothesis variance arising from the differences among the posterior means under the three hypotheses. Under the changepoint hypothesis, the within-hypothesis variance is zero, because observation $o_t$ is assumed to uniquely determine the new latent mean.